\documentstyle[floats,tighten,preprint,eqsecnum,prd,aps,graphicx]{revtex}

\newlength{\PlotHeight}
\begin{document}


\preprint{\parbox[t]{15em}{\raggedleft
FERMILAB-PUB-01/317-T \\ hep-ph/0110253\\[2.0em]}}
\draft

\title{Lattice calculation of the zero-recoil form factor of 
$\bar{B}\to D^*l\bar{\nu}$:
toward a model independent determination of~$|V_{cb}|$}

\author{Shoji~Hashimoto,$^1$
    Andreas~S.~Kronfeld,$^2$
      Paul~B.~Mackenzie,$^2$
     Sin\'{e}ad~M.~Ryan,$^3$ and
         James~N.~Simone$^2$}

\address{%
$^1$High Energy Accelerator Research Organization (KEK), 
	Tsukuba~305-0801, Japan \\
$^2$Fermi National Accelerator Laboratory,
	Batavia, Illinois 60510 \\
$^3$School of Mathematics, Trinity College,
	Dublin~2, Ireland}

\date{\today}
\maketitle

\begin{abstract}
We develop a new method in lattice QCD to calculate the form factor
${\cal F}_{B\to D^*}(1)$ at zero recoil.
This is the main theoretical ingredient needed to determine~$|V_{cb}|$
from the exclusive decay $\bar{B}\to D^*l\bar{\nu}$.
We introduce three ratios, in which most of statistical and systematic 
error cancels, making a precise calculation possible.
We fit the heavy-quark mass dependence directly, and extract the
$1/m_Q^2$ and three of the four $1/m_Q^3$ corrections in the heavy-quark
expansion.
In this paper we show how the method works in the quenched 
approximation, obtaining
${\cal F}_{B\to D^*}(1)=0.913%
		  ^{+0.024}_{-0.017}
		\pm 0.016
		{}^{+0.003}_{-0.014}
		{}^{+0.000}_{-0.016}
		{}^{+0.006}_{-0.014}$
where the uncertainties come, respectively, from
statistics and fitting,
matching lattice gauge theory to QCD,
lattice spacing dependence,
light quark mass effects,
and the quenched approximation.
We also discuss how to reduce these uncertainties and, thus, to
obtain a model-independent determination of~$|V_{cb}|$.
\end{abstract}




\section{Introduction}
\label{sec:intro}

In flavor physics the Cabbibo-Kobayashi-Maskawa (CKM) matrix
element~$V_{cb}$ plays an important role.
Much of the phenomenology of $CP$ violation centers around the
unitarity triangle, and a precise value of~$|V_{cb}|$ is needed to locate
the triangle's apex in the complex plane.
As a fundamental parameter of the Standard Model, $V_{cb}$ sometimes
appears in unexpected places.
For example, the Standard Model prediction of the $K^0$-$\bar{K}^0$ 
mixing parameter $\epsilon_K$ is very sensitive 
to~$|V_{cb}|$~\cite{Rosner:1998gb}.

The determination of~$|V_{cb}|$ is made through inclusive and exclusive
semileptonic~$B$ decays, but at present both methods are limited by
theoretical uncertainties.
The inclusive method requires a reliable calculation of the total
semileptonic decay rate of the $B$ meson, which can be done using the
heavy quark expansion~\cite{Ball:1995wa,Bigi:1997fj}.
Ultimately this method is limited by the breakdown of local 
quark-hadron duality, which is difficult to estimate.
The exclusive method, on the other hand, requires a theoretical 
calculation of the form factor~${\cal F}_{B\to D^*}$ of
$\bar{B}\to D^*l\bar{\nu}$ decay.
In this paper we take a step towards reducing the uncertainty in
the exclusive method, by devising a precise method to compute
the form factor at zero recoil in lattice~QCD.

The differential rate for the semileptonic decay 
$\bar{B}\to D^*l\bar{\nu}_l$ is given by
\begin{equation}
	\frac{d\Gamma}{dw} =
		\frac{G_F^2}{4\pi^3} m_{D^*}^3 (m_B-m_{D^*})^2
		\sqrt{w^2-1} {\cal G}(w) 
		|V_{cb}|^2 |{\cal F}_{B\to D^*}(w)|^2,
	\label{eq:dGamma/dw}
\end{equation}
where $w=v'\cdot v$ is the velocity transfer from the initial state
(with velocity~$v$) to the final state (with velocity~$v'$).
The velocity transfer is related to the momentum~$q$ transferred to the
leptons by $q^2=m_B^2 - 2wm_Bm_{D^*} + m_{D^*}^2$,
and it lies in the range $1\leq w < (m_B^2+ m_{D^*}^2)/2m_Bm_{D^*}$.
The function
\begin{equation}
	{\cal G}(w) = \frac{w+1}{12} \left( 5w + 1 +
		\frac{8w(w-1)m_Bm_{D^*}}{(m_B-m_{D^*})^2} \right)
	\label{eq:Gw}
\end{equation}
has a kinematic origin, with ${\cal G}(1)=1$.
Thus, given the form factor ${\cal F}_{B\to D^*}(w)$, one can use the
measured decay rate to determine~$|V_{cb}|$.

One makes use of the zero-recoil point $w=1$, even though the
phase-space factor $\sqrt{w^2-1}$ suppresses the event rate, because
then theoretical uncertainties are under better control.
For $w>1$, ${\cal F}_{B\to D^*}(w)$ is a linear combination of several
form factors of $\bar{B}\to D^*$ transitions mediated by the vector and axial
vector currents.
At zero recoil, however,
\begin{equation}
	{\cal F}_{B\to D^*}(1) = h_{A_1}(1),
	\label{eq:F=hA1}
\end{equation}
where $h_{A_1}$ is a form factor of the axial vector
current~${\cal A}^\mu$, namely,
\begin{equation}
	\langle D^*(v) | {\cal A}^\mu | \bar{B}(v) \rangle =
		i \sqrt{2m_B\,2m_{D^*}} \,\overline{\epsilon'}^\mu h_{A_1}(1).
	\label{eq:DAB}
\end{equation}
More importantly, heavy-quark symmetry plays an essential role in
constraining~$h_{A_1}(1)$, leading to the simple heavy quark
expansion~\cite{Falk:1993wt,Mannel:1994kv}
\begin{equation}
	h_{A_1}(1) = \eta_A \left[1 -
		\frac{\ell_V}{(2m_c)^2} +
		\frac{2\ell_A}{2m_c\,2m_b} -
		\frac{\ell_P}{(2m_b)^2} \right],
	\label{eq:hA1HQE}
\end{equation}
including all terms of order $1/m_Q^2$.
In Eq.~(\ref{eq:hA1HQE}), $\eta_A$ is a short-distance radiative
correction, which is known at the two-loop
level~\cite{Czarnecki:1996gu,Czarnecki:1997cf},
and the $\ell$s are long-distance matrix elements of the heavy-quark
effective theory~(HQET).%
\footnote{In the HQET literature, the $\ell$s are often called
``hadronic parameters'', because they are viewed as incalculable.
In a QCD context, however, the are not free parameters, but calculable
matrix elements.}
Heavy-quark symmetry normalizes the leading term inside the bracket to
unity~\cite{Isgur:1989vq} and, moreover, forbids terms
of order $1/m_Q$~\cite{Luke:1990eg}.
The~$1/m_Q^2$ corrections are formally
small---$(\bar{\Lambda}/2m_c)^2\sim4\%$---but one would like to reach
better precision on~$|V_{cb}|$, so these terms cannot be neglected.

There have been mainly two different methods used to estimate the
$1/m_Q^2$ terms in Eq.~(\ref{eq:hA1HQE}), but neither has achieved a
model independent calculation.
One involves using a quark model~\cite{Falk:1993wt,Neubert:1994vy}
to estimate the~$\ell$s.
The other employs the zero-recoil sum rule~\cite{Shifman:1995jh}.
Although based on a rigorous upper bound~\cite{Bigi:1995ga}, to make
a prediction of ${\cal F}_{B\to D^*}(1)$ this approach requires an
assumption on the effects of higher excited states in the sum rule.
Thus---just as with quark models---it is difficult to estimate, let
alone reduce, the uncertainty associated with the estimate.

In this paper we take a step towards reducing the theoretical 
uncertainty by using lattice QCD to
calculate~$h_{A_1}(1)={\cal F}_{B\to D^*}(1)$.
Lattice QCD is, in principle, model independent, although here we
work in the quenched approximation.
The quenched approximation is not less rigorous than the methods used
in Refs.~\cite{Neubert:1994vy,Shifman:1995jh}.
From our point of view, however, the main advantage of the quenched
approximation is that it allows us to learn how to control and estimate
all other lattice uncertainties.
With a proven technique, it is conceptually straightforward, if
computationally demanding, to carry out a calculation in full~QCD.

Until now three obstacles prevented even quenched lattice calculations
of $h_{A_1}(1)$ to the needed precision.
First, a direct Monte Carlo calculation of the matrix element in
Eq.~(\ref{eq:DAB}) suffers from a statistical error that is too large to
be interesting.
Second, the normalization of the lattice axial vector current was
uncertain, being limited by a poorly converging perturbation series.
Finally, early works~\cite{Booth:1994zb} used \emph{ad hoc} methods for
heavy quarks on the lattice, which entailed a poorly controlled
extrapolation in the heavy quark mass.
We have devised methods to circumvent all three obstacles.
The first two are handled with certain double ratios of correlation
functions, in which the bulk of statistical and systematic uncertainties
cancel~\cite{Hashimoto:2000yp}.
The third obstacle---the problem of heavy-quark lattice artifacts---is
overcome by using a systematic method for treating heavy quarks on
the lattice, based on Wilson fermions~\cite{El-Khadra:1997mp}.
This obstacle could also be overcome using lattice
NRQCD~\cite{Thacker:1991bm},
as in the work of Hein \emph{et al}.~\cite{Hein:2000se}.

In our work~\cite{Hashimoto:2000yp} on the form factor~$h_+(1)$ in
the decay $\bar{B}\to Dl\bar{\nu}$ at zero recoil, a central role was
played by the double ratio of matrix elements
\begin{equation}
	{\cal R}_+ =
	\frac{
	\langle     D   |\bar{c}\gamma^4b|  \bar{B}\rangle
	\langle\bar{B}  |\bar{b}\gamma^4c|       D \rangle
	}{
	\langle     D   |\bar{c}\gamma^4c|       D \rangle
	\langle\bar{B}  |\bar{b}\gamma^4b|  \bar{B}\rangle
	} = |h_+(1)|^2,
	\label{eq:contR+}
\end{equation}
where
\begin{equation}
	\langle D(v) | {\cal V}^\mu | \bar{B}(v) \rangle =
		i \sqrt{2m_B\,2m_D} \, v^\mu h_+(1).
	\label{eq:DVB}
\end{equation}
In Ref.~\cite{Hashimoto:2000yp} we studied the heavy-quark mass
dependence of~$h_+(1)$, using a fit to obtain the $1/m_Q^2$ and
$1/m_Q^3$ corrections.
In this work we employ this double ratio and two similar ones.
The first additional double ratio is
\begin{equation}
	{\cal R}_1 =
	\frac{
	\langle     D ^*|\bar{c}\gamma^4b|\bar{B}^*\rangle
	\langle\bar{B}^*|\bar{b}\gamma^4c|     D^* \rangle
	}{
	\langle     D ^*|\bar{c}\gamma^4c|     D^* \rangle
	\langle\bar{B}^*|\bar{b}\gamma^4b|\bar{B}^*\rangle
	} = |h_1(1)|^2,
	\label{eq:contR1}
\end{equation}
where the pseudoscalar mesons $\bar{B}$ and $D$, and their form
factor~$h_+(1)$, are replaced with the vector mesons
$\bar{B}^*$ and $D^*$, and their form factor~$h_1(1)$:
\begin{equation}
	\langle D^*(v) | {\cal V}^\mu | \bar{B}^*(v) \rangle =
		i \sqrt{2m_{B^*}\,2m_{D^*}} \,
		\overline{\epsilon'}\cdot\epsilon \, v^\mu h_1(1).
	\label{eq:D*VB*}
\end{equation}
The second additional double ratio is
\begin{equation}
	{\cal R}_{A_1} =
	\frac{
	\langle     D ^*|\bar{c}\gamma_j\gamma_5b|\bar{B}  \rangle
	\langle\bar{B}^*|\bar{b}\gamma_j\gamma_5c|     D   \rangle
	}{
	\langle     D ^*|\bar{c}\gamma_j\gamma_5c|     D   \rangle
	\langle\bar{B}^*|\bar{b}\gamma_j\gamma_5b|\bar{B}  \rangle
	} = 
	\frac{h_{A_1}^{\bar{B}\to D^*}(1) h_{A_1}^{D\to \bar{B}^*}(1)}%
	     {h_{A_1}^{D\to D^*}(1) h_{A_1}^{\bar{B}\to \bar{B}^*}(1)} \equiv
	|\check{h}_{A_1}(1)|^2,
	\label{eq:contRA1}
\end{equation}
where the axial vector current mediates pseudoscalar-to-vector
transitions, leading to a double ratio of the form factor~$h_{A_1}$.
As stressed in Ref.~\cite{Hashimoto:2000yp}, the double ratios
overcome two of the obstacles in the lattice calculation, because
numerator and denominator are so similar.
Statistical fluctuations in the numerator and denominator are very
highly correlated and largely cancel in the ratio.
Also, most of the normalization uncertainty in the lattice currents
cancels, leaving only a residual normalization factor that can
be computed reliably in perturbation theory~\cite{Kronfeld:1999tk}.
Indeed, all uncertainties scale as ${\cal R}-1$, rather
than as~${\cal R}$.

Note that the double ratio~${\cal R}_{A_1}$ does not yield
the desired form factor~$h^{\bar{B}\to D^*}_{A_1}$, but instead the
combination~$\check{h}_{A_1}$, which is itself a double ratio of
form factors.
One can, however, extract $h_{A_1}(1)$ from the three double ratios
${\cal R}_+$, ${\cal R}_1$, and ${\cal R}_{A_1}$, at least to the
order in the heavy-quark expansion given in Eq.~(\ref{eq:hA1HQE}).
This possibility follows from the heavy quark expansions for $h_+(1)$
and $h_1(1)$~\cite{Falk:1993wt,Mannel:1994kv},
\begin{eqnarray}
	h_+(1) & = & \eta_V \left[ 1 - \ell_P 
		\left( \frac{1}{2m_c} - \frac{1}{2m_b} \right)^2 \right],
	\label{eq:h+HQE} \\
	h_1(1) & = & \eta_V \left[ 1 - \ell_V 
		\left( \frac{1}{2m_c} - \frac{1}{2m_b} \right)^2  \right],
	\label{eq:h1HQE}
\end{eqnarray}
and comparing to Eq.~(\ref{eq:hA1HQE}).
In $h_+(1)$ and $h_1(1)$ the absence of terms of order
$1/m_Q$~\cite{Luke:1990eg} is easily understood, because charge
conservation requires $h_+(1)=h_1(1)=1$ when $m_c=m_b$, and because
the matrix elements defining them are symmetric under the interchange
$m_c\leftrightarrow m_b$.
Similarly, the heavy-quark expansion of the form factor
ratio~$\check{h}_{A_1}(1)$, obtained from ${\cal R}_{A_1}$, is
\begin{equation}
	\check{h}_{A_1}(1) = \check{\eta}_A \left[ 1 - \ell_A 
		\left( \frac{1}{2m_c} - \frac{1}{2m_b} \right)^2 \right],
	\label{eq:checkhA1HQE}
\end{equation}
which follows immediately from Eq.~(\ref{eq:hA1HQE}), defining
$\check{\eta}_A^2=%
\eta_{A^{cb}} \eta_{A^{bc}}/\eta_{A^{cc}} \eta_{A^{bb}}$.
Hence, by varying the heavy quark masses in the lattice calculation of
the double ratios ${\cal R}_+$, ${\cal R}_1$, and ${\cal R}_{A_1}$,
one can extract $\ell_P$, $\ell_V$, and $\ell_A$, respectively.
Then, $h_{A_1}(1)={\cal F}_{B\to D^*}(1)$ can be reconstituted
through Eq.~(\ref{eq:hA1HQE}).

A key to this method is that heavy-quark symmetry
requires the quantities~$\ell_P$ and~$\ell_V$ to appear in
Eq.~(\ref{eq:hA1HQE}), as well as in Eqs.~(\ref{eq:h+HQE})
and~(\ref{eq:h1HQE})~\cite{Falk:1993wt,Mannel:1994kv}.
A~simple argument explains why.
For each form factor there are three possible terms at order
$1/m_Q^2$---$1/m_c^2$, $1/m_b^2$, and $1/m_c m_b$---and each multiplies
an HQET matrix element.
For $h_+(1)$ and $h_1(1)$ the particular form of the expansions is
restricted by the $b\leftrightarrow c$ interchange symmetry, so only
one HQET matrix element can appear in each case: $\ell_P$ for $h_+(1)$
and $\ell_V$ for~$h_1(1)$.
Interchange symmetry does not apply to the $\bar{B}\to D^*$ transition,
however, so three HQET matrix elements are needed in the expansion of
$h_{A_1}(1)$, Eq.~(\ref{eq:hA1HQE}).
Two of them, however, coincide with $\ell_P$ and~$\ell_V$.
If one flips the spin of the charmed quark in the $\bar{B}\to D$ transition
in Eq.~(\ref{eq:DVB}), one obtains the $\bar{B}\to D^*$ transition in
Eq.~(\ref{eq:DAB}), and in the limit of infinite charmed quark mass
the matrix elements are identical, by heavy-quark spin symmetry.
Consequently, the $1/m_b^2$ term in Eq.~(\ref{eq:hA1HQE}) must be the
same as that in Eq.~(\ref{eq:h+HQE}), namely $\ell_P/(2m_b)^2$.
The same logic applied to the $b$ quark's spin, starting from
the $\bar{B}^*\to D^*$ transition in Eq.~(\ref{eq:D*VB*}), implies that
the~$1/m_c^2$ term in Eqs.~(\ref{eq:hA1HQE}) and~(\ref{eq:h1HQE})
must be the same, namely~$\ell_V/(2m_c)^2$.

At order $1/m_Q^3$ there are, in general, four terms for each form
factor.
In Sec.~\ref{sec:results} we show how the same kind of reasoning can
be used to extract three of the four terms from the $1/m_Q^3$ behavior
of the three double ratios.
Including these corrections not only reduces the systematic error of the
heavy quark expansion, but also reduces our statistical error, because
fitted values for the quadratic and cubic terms are correlated.

In the remainder of this paper we describe the details of our lattice
calculation of ${\cal F}_{B\to D^*}(1)=h_{A_1}(1)$, as sketched
above.
Discretization effects are studied by repeating the analysis at
three different lattice spacings.
The dependence on the light quark mass is expected to be small, which we
are able to verify.
After a thorough investigation of systematic uncertainties, we obtain
\begin{equation}
	{\cal F}_{B\to D^*}(1) = 0.913
		^{+0.024}_{-0.017}
		\pm 0.016
		{}^{+0.003}_{-0.014}
		{}^{+0.000}_{-0.016}
		{}^{+0.006}_{-0.014}
	\label{eq:result}
\end{equation}
where the uncertainties come, respectively, from
statistics and fitting,
matching lattice gauge theory and HQET to QCD,
lattice spacing dependence,
light quark mass effects,
and the quenched approximation.
A~preliminary report of this calculation based on our coarsest lattice
appeared in Ref.~\cite{Simone:2000yn}, reporting
${\cal F}_{B\to D^*}(1)=0.935\pm0.022^{+0.023}_{-0.024}$.
The change comes mostly from the results on two finer lattices,
partly from some secondary changes in the analysis, and
partly from the inclusion of some contributions of order~$1/m_Q^3$.
Clearly, these central values are indistinguishable within the error
bars.

The paper is organized as follows.
In Sec.~\ref{sec:bckgnd} we discuss how to combine heavy-quark theory
and lattice gauge theory to calculate the needed matrix elements;
in particular, we review how we are able to extract the $1/m_Q^2$
corrections~\cite{Kronfeld:2000ck}.
Section~\ref{sec:bckgnd} is fairly general and much of it also applies
to lattice~NRQCD.
Specific details of our numerical work are given in 
Sec.~\ref{sec:Lattice_calculation}, including input parameters and the
basic outputs.
The ``Fermilab'' method for heavy quarks~\cite{El-Khadra:1997mp}
requires matching the short-distance behavior of lattice gauge theory to
QCD, which is discussed in Sec.~\ref{sec:PT}.
Section~\ref{sec:results} shows a key feature of our analysis, namely
the direct fitting of the heavy-quark mass dependence to obtain the
power corrections in Eq.~(\ref{eq:hA1HQE}).
A~detailed discussion of the systematic uncertainties is in
Sec.~\ref{sec:errors}.
Our result, Eq.~(\ref{eq:result}), is compared to other methods in
Sec.~\ref{sec:comparison}.
Section~\ref{sec:conclusion} contains some concluding remarks.

\section{Continuum and lattice matrix elements}
\label{sec:bckgnd}

In this section we discuss how to obtain continuum-QCD, heavy-quark
observables from lattice gauge theory.
Discretization effects of the heavy quarks are a special concern, so
they are discussed in detail in this section.
For the light spectator quark we use well-known methods, and we provide
details in Sec.~\ref{sec:Lattice_calculation}.

Discretization effects of the heavy quarks can be controlled by matching
the lattice theory to HQET~\cite{Kronfeld:2000ck}.
This is possible whether one discretizes the NRQCD effective
Lagrangian~\cite{Thacker:1991bm},
or one employs the non-relativistic interpretation of Wilson
fermions~\cite{El-Khadra:1997mp}.
In either case, on-shell lattice matrix elements can be described by a
version of (continuum) HQET, with effective Lagrangian (in the rest
frame)
\begin{equation}
	{\cal L}_{\text{HQET}} = m_1\bar{h}_vh_v +
		\frac{\bar{h}_v\bbox{D}^2h_v}{2m_2} +
		\frac{\bar{h}_v\,i\bbox{\Sigma}\cdot\bbox{B}\,h_v}{2m_{\cal B}}
		+ \cdots,
	\label{eq:LHQET}
\end{equation}
where $h_v$ is the heavy-quark field of~HQET, and $\bbox{B}$ is the
chromomagnetic field.
The ``masses'' $m_1$, $m_2$, and $m_{\cal B}$ are short-distance
coefficients; they depend on the bare couplings of the lattice action,
including the gauge coupling.
Matrix elements are completely independent
of~$m_1$~\cite{Kronfeld:2000ck}, so the important coefficients are
$m_2$ and~$m_{\cal B}$.
The lattice NRQCD action has bare parameters that correspond directly
to $m_2$ and~$m_{\cal B}$.
With Wilson fermions one must use the Sheikholeslami-Wohlert (SW)
action~\cite{Sheikholeslami:1985ij}, and adjust $m_0$ and
$c_{\text{SW}}$ to tune $m_2$ and~$m_{\cal B}$.
In practice, we tune $m_2$ non-perturbatively, using the heavy-light and
quarkonium spectra, and $m_{\cal B}$ with the estimate of
tadpole-improved, tree-level perturbation theory~\cite{Lepage:1993xa}.
There are also terms of order $1/m_Q^2$ in the effective
Lagrangian~${\cal L}_{\text{HQET}}$, but they do not influence the
double ratios, as discussed further below.

In this paper we use lattice currents that are constructed as in
Ref.~\cite{El-Khadra:1997mp}.
(An analogous set of currents can be constructed for lattice
NRQCD~\cite{Boyle:2000fi}.)
We distinguish the lattice currents~$V^\mu$ and~$A^\mu$ from their
continuum counterparts~${\cal V}^\mu$ and~${\cal A}^\mu$.
We define
\begin{eqnarray}
	V^\mu & = & \sqrt{Z_{V^{cc}} Z_{V^{bb}}}
		\bar{\Psi}_ci\gamma^\mu \Psi_b
	\label{eq:Vlat} \\
	A^\mu & = & \sqrt{Z_{V^{cc}} Z_{V^{bb}}}
		\bar{\Psi}_ci\gamma^\mu \gamma_5 \Psi_b
	\label{eq:Alat}
\end{eqnarray}
where the rotated field~\cite{El-Khadra:1997mp}
\begin{equation}
	\Psi_q = \left[1 + ad_1
	\bbox{\gamma}\cdot\bbox{D}_{\text{lat}}\right] \psi_q,
	\label{eq:rotate}
\end{equation}
and $\psi_q$ is the lattice quark field ($q=c,\;b$) in the SW action.
Here $\bbox{D}_{\text{lat}}$ is the symmetric, nearest-neighbor, 
covariant difference operator.
In Eqs.~(\ref{eq:Vlat}) and~(\ref{eq:Alat}) the factors $Z_{V^{qq}}$,
$q=c,b$, normalize the flavor-conserving vector currents.
Because for massive quarks only~$Z_V$ can be computed 
non-perturbatively, we choose to put $Z_V$ into the definition of
the axial current~$A^\mu$.
In the work reported in this paper, we do not need to compute the
factor~$\sqrt{Z_{V^{cc}} Z_{V^{bb}}}$, because it cancels in the
double ratios.

Matching the current~$V^\mu$ to HQET requires further short-distance
coefficients:
\begin{eqnarray}
	V^\mu & \doteq & C^{\text{lat}}_{V_\parallel} v^\mu \bar{c}_vb_v
		- \frac{B^{\text{lat}}_{Vc} \bar{c}_v
			\loarrow{\kern+0.1em /\kern-0.65em D}_\perp
			i\gamma^\mu_\perp b_v}{2m_{3c}}
		- \frac{B^{\text{lat}}_{Vb} \bar{c}_vi\gamma^\mu_\perp
			{\kern+0.1em /\kern-0.65em D}_\perp b_v}{2m_{3b}} + \cdots,
	\label{eq:VlatHQET} \\
	A^\mu & \doteq &
		C^{\text{lat}}_{A_\perp} \bar{c}_vi\gamma^\mu_\perp\gamma_5 b_v
		+ \frac{B^{\text{lat}}_{Ac} v^\mu \bar{c}_v
			\loarrow{\kern+0.1em /\kern-0.65em D}_\perp
			\gamma_5 b_v}{2m_{3c}}
		- \frac{B^{\text{lat}}_{Ab} v^\mu \bar{c}_v {\gamma_5}
			{\kern+0.1em /\kern-0.65em D}_\perp b_v}{2m_{3b}} + \cdots,
	\label{eq:AlatHQET}
\end{eqnarray}
where the symbol $\doteq$ implies equality of matrix elements, and
$b_v$ and $c_v$ are HQET fields for the bottom and charmed quarks.
At the tree level the short-distance coefficients 
$C^{\text{lat}}_{V_\parallel}$, $C^{\text{lat}}_{A_\perp}$, and
$B^{\text{lat}}_{hJ}$ all equal~one.
The free parameter $d_1$ in Eq.~(\ref{eq:rotate}) can be adjusted
to tune $1/m_{3Q}$ to $1/m_Q$.
In the present calculations, we adjust~$d_1$ with the estimate of
tadpole-improved, tree-level perturbation theory, as explained in 
Ref.~\cite{El-Khadra:1997mp}.
Further dimension-four operators, whose coefficients vanish at the tree
level, are omitted from the right-hand sides of Eqs.~(\ref{eq:VlatHQET})
and~(\ref{eq:AlatHQET}); they are listed in Ref.~\cite{Kronfeld:1999tk}.

The description in Eqs.~(\ref{eq:VlatHQET}) and~(\ref{eq:AlatHQET}) is in
complete analogy with that for the continuum currents, namely,
\begin{eqnarray}
	{\cal V}^\mu & \doteq & C_{V_\parallel} v^\mu \bar{c}_vb_v
		- \frac{B_{Vc} \bar{c}_v
			\loarrow{\kern+0.1em /\kern-0.65em D}_\perp
			i\gamma^\mu_\perp b_v}{2m_{c}}
		- \frac{B_{Vb} \bar{c}_vi\gamma^\mu_\perp
			{\kern+0.1em /\kern-0.65em D}_\perp b_v}{2m_{b}} + \cdots,
	\label{eq:VcontHQET} \\
	{\cal A}^\mu & \doteq &
		C_{A_\perp} \bar{c}_vi\gamma^\mu_\perp\gamma_5 b_v
		+ \frac{B_{Ac} v^\mu \bar{c}_v
			\loarrow{\kern+0.1em /\kern-0.65em D}_\perp
			\gamma_5 b_v}{2m_{c}}
		- \frac{B_{Ab} v^\mu \bar{c}_v {\gamma_5}
			{\kern+0.1em /\kern-0.65em D}_\perp b_v}{2m_{b}} + \cdots.
	\label{eq:AcontHQET}
\end{eqnarray}
The radiative corrections to the short-distance coefficients in
Eqs.~(\ref{eq:VlatHQET})  and~(\ref{eq:AlatHQET}) differ from those in
Eqs.~(\ref{eq:VcontHQET}) and~(\ref{eq:AcontHQET}), because the
lattice modifies the physics at short distances.
On the other hand, the HQET operators are the same throughout.

There are also terms of order $1/m_Q^2$ in the effective currents on the
right-hand sides of Eqs.~(\ref{eq:VlatHQET})--(\ref{eq:AcontHQET}),
although for brevity they are not written out.
The most important operator in each case is
\begin{eqnarray}
	V_{(1,1)}^\mu & = &
		\frac{\bar{c}_v \loarrow{\kern+0.1em /\kern-0.65em D}_\perp
		v^\mu {\kern+0.1em /\kern-0.65em D}_\perp b_v}{2m_{3c}\,2m_{3b}},
		\label{eq:V1,1} \\
	A_{(1,1)}^\mu & = &
		\frac{\bar{c}_v \loarrow{\kern+0.1em /\kern-0.65em D}_\perp
		i\gamma^\mu_\perp \gamma_5
		{\kern+0.1em /\kern-0.65em D}_\perp b_v}{2m_{3c}\,2m_{3b}}.
		\label{eq:A1,1}
\end{eqnarray}
As the notation suggests, both these currents are correctly normalized
at the tree level when $d_1$ is adjusted so that $m_{3Q}=m_Q$, as above.
In addition to these $1/m_cm_b$ currents, there are currents of
order~$1/m_c^2$ and~$1/m_b^2$.
Although the latter contribute to the individual matrix elements
$\langle D^{(*)}|J^\mu|B^{(*)}\rangle$, their contributions drop out of 
the double ratios.

In the foregoing discussion, most corrections of order $1/m_Q^2$ have 
been handled only in a cursory way.
Since we aim for the $1/m_Q^2$ corrections to the double ratios we must,
however, discuss how these contributions are incorporated, when the
lattice action and currents are constructed and normalized along the
lines given above.
The HQET description of matrix elements reveals several sources of
such contributions~\cite{Falk:1993wt,Mannel:1994kv,Kronfeld:2000ck}:
\begin{enumerate}
	\item double insertions of the $1/m_Q$ terms in the effective
	Lagrangian~${\cal L}_{\text{HQET}}$;
	\item single insertions of the $1/m_Q$ terms in the effective
	Lagrangian into matrix elements of the $1/m_Q$ terms in the
	effective HQET currents;
	\item single insertions of genuine $1/m_Q^2$ terms in the effective
	Lagrangian;
	\item matrix elements of genuine $1/m_Q^2$ terms in the effective
	HQET currents.
\end{enumerate}
The first set of contributions is correctly normalized at the same level
of accuracy as the $1/m_Q$ terms of the action.
The second set makes no contribution to zero recoil matrix elements
whatsoever~\cite{Kronfeld:2000ck}.
The third set also makes no contribution at zero recoil, because the
leading terms in Eqs.~(\ref{eq:VlatHQET}) and (\ref{eq:AlatHQET})
are Noether currents of the heavy-quark symmetries and, as in the
proof of Luke's theorem, first corrections to Noether currents
vanish~\cite{Ademollo:1964sr,Kronfeld:2000ck}.

One is left with the last set, which \emph{does} contribute to the
matrix elements defining the form factors.
The HQET matrix elements of all dimension-five currents can be reduced
to $\lambda_1$ and $\lambda_2$, which appear in the heavy-quark
expansion of the mass~\cite{Falk:1993wt}.
In the double ratios, however, the following cancellation
(schematically) takes place~\cite{Kronfeld:2000ck}:
\begin{equation}
	\frac{
	[1 - \lambda(X_b/m_b^2 - 1/m_cm_b + X_c/m_c^2)]^2
	}{
	[1 - \lambda(2X_c - 1)/m_c^2] [1 - \lambda(2X_b - 1)/m_b^2]
	} =
	1 - \lambda\left(\frac{1}{m_c} - \frac{1}{m_b}\right)^2,
\end{equation}
where $\lambda$ is proportional to $\lambda_1$ or $\lambda_2$,
and $X_Q/m_Q^2$ indicates \emph{incorrect} normalization,
while $1/m_Qm_{Q'}$ indicates \emph{correct}   normalization.
In practice, the ``correctly normalized'' terms are normalized only
at the tree level.
Nevertheless, the double ratios suffer from uncertainties only of order
$\alpha_s(\bar{\Lambda}/m_Q)^2$, even though the action is matched
only at the $1/m_Q$ level and the currents are matched only at the
$1/m_cm_b$ level.

Once one is content to neglect corrections of order
$\alpha_s(\bar{\Lambda}/m_Q)^2$, it is easy to obtain
the continuum normalization of the lattice currents.
By comparing the heavy-quark expansions for $V^\mu$ and $A^\mu$ to those
for ${\cal V}^\mu$ and ${\cal A}^\mu$, one sees that
\begin{eqnarray}
	{\cal V}^\mu_{cb} & \doteq & \rho_{V^{cb}} V^\mu_{cb},
	\label{eq:VrhoV}  \\
	{\cal A}^\mu_{cb} & \doteq & \rho_{A^{cb}} A^\mu_{cb},
	\label{eq:ArhoA}
\end{eqnarray}
apart from discretization effects discussed above.
The $\rho$ factors are
\begin{eqnarray}
	\rho_{V^{cb}} & = & C_{V_\parallel}/C^{\text{lat}}_{V_\parallel},
	\label{eq:rhoVCC}  \\
	\rho_{A^{cb}} & = & C_{A_\perp}/C^{\text{lat}}_{A_\perp},
	\label{eq:rhoACC}
\end{eqnarray}
and they are known at the one-loop level~\cite{Kronfeld:1999tk}. 

The matrix elements are obtained from three-point correlation functions.
For the zero-recoil $B\to D$, $B^*\to D^*$ and $B\to D^*$ transitions
the three-point function are, respectively,
\begin{eqnarray}
	C^{B\to D}(t_f,t_s,t_i)     & = & \sum_{\bbox{x},\bbox{y}}
	\langle 0| {\cal O}_D(\bbox{x},t_f)
	\bar{\Psi}_c\gamma_4\Psi_b(\bbox{y},t_s)
	{\cal O}^\dagger_B(\bbox{0}, t_i)|0\rangle ,
	\label{eq:3ptB2D} \\
	C^{B^*\to D^*}(t_f,t_s,t_i) & = & \sum_{\bbox{x},\bbox{y}}
	\langle 0| {\cal O}_{D^*}(\bbox{x},t_f)
	\bar{\Psi}_c\gamma_4\Psi_b(\bbox{y},t_s)
	{\cal O}^\dagger_{B^*}(\bbox{0}, t_i)|0\rangle ,
	\label{eq:3ptB*2D*} \\
	C^{B\to D^*}(t_f,t_s,t_i)   & = & \sum_{\bbox{x},\bbox{y}}
	\langle 0| {\cal O}_{D^*}(\bbox{x},t_f)
	\bar{\Psi}_c\gamma_j\gamma_5\Psi_b(\bbox{y},t_s)
	{\cal O}^\dagger_B(\bbox{0}, t_i)|0\rangle ,
	\label{eq:3ptB2D*}
\end{eqnarray}
where ${\cal O}_{B^{(*)}}$ and ${\cal O}_{D^{(*)}}$ are interpolating
operators for the $B^{(*)}$ and $D^{(*)}$ mesons.
In $C^{B^*\to D^*}$ the spins of the vector mesons are parallel, and
in $C^{B\to D^*}$   the spin  of the $D^*$ lies in the $j$~direction.
These correlation functions are calculated by a Monte Carlo method,
as usual in lattice~QCD.
In the limit of large time separations, the correlation functions become
\begin{eqnarray}
	C^{B\to D}(t_f,t_s,t_i)     & = &
		{\cal Z}_D^{1/2}     {\cal Z}_B^{1/2}
	\frac{\langle D| \bar{\Psi}_c\gamma_4\Psi_b |B\rangle}%
	{\sqrt{2m_D}\hfil\sqrt{2m_B}}
		e^{-m_B(t_s-t_i)}    e^{-m_D(t_f-t_s)} + \cdots,
	\label{three-ptB2D} \\
	C^{B^*\to D^*}(t_f,t_s,t_i) & = & 
		{\cal Z}_{D^*}^{1/2} {\cal Z}_{B^*}^{1/2}
	\frac{\langle D^*| \bar{\Psi}_c\gamma_4\Psi_b |B^*\rangle}%
	{\sqrt{2m_{D^*}}\hfil\sqrt{2m_{B^*}}}
		e^{-m_{B^*}(t_s-t_i)}e^{-m_{D^*}(t_f-t_s)} + \cdots,
	\label{three-ptB*2D*} \\
	C^{B\to D^*}(t_f,t_s,t_i)   & = & 
		{\cal Z}_{D^*}^{1/2} {\cal Z}_B^{1/2}
	\frac{\langle D^*| \bar{\Psi}_c\gamma_j\gamma_5\Psi_b |B\rangle}%
	{\sqrt{2m_{D^*}}\hfil\sqrt{2m_B}}
		e^{-m_B(t_s-t_i)}    e^{-m_{D^*}(t_f-t_s)} + \cdots,
	\label{three-ptB2D*}
\end{eqnarray}
where $m_{B^{(*)}}$ and $m_{D^{(*)}}$ are the masses of the $B^{(*)}$
and $D^{(*)}$ mesons.
The normalization factors $\sqrt{{\cal Z}_{H^{(*)}}/2m_{H^{(*)}}}$
are conventional; they cancel when forming the double ratios,
so we do not need them.
The correlation functions defined in
Eqs.~(\ref{eq:3ptB2D})--(\ref{eq:3ptB2D*})
are the only objects needed from the Monte Carlo.
In practice we hold $t_i=0$ and $t_f=T/2$ fixed and vary
$t_s$ over the range for which the lowest-lying states
dominate the correlation functions, as is needed for
Eqs.~(\ref{three-ptB2D})--(\ref{three-ptB2D*}) to~hold.
($T=N_Ta$ is the temporal length of the lattice.)

From the correlation functions we form the following double ratios
\begin{eqnarray}
	R_+(t) & = & 
	\frac{C^{B  \to D  }(0,t,T/2) \; C^{D  \to B  }(0,t,T/2)}%
	     {C^{D  \to D  }(0,t,T/2) \; C^{B  \to B  }(0,t,T/2)},
	\label{eq:R+def} \\
	R_1(t) & = & 
	\frac{C^{B^*\to D^*}(0,t,T/2) \; C^{D^*\to B^*}(0,t,T/2)}%
	     {C^{D^*\to D^*}(0,t,T/2) \; C^{B^*\to B^*}(0,t,T/2)},
	\label{eq:R1def} \\
	R_{A_1}(t) & = & 
	\frac{C^{B  \to D^*}(0,t,T/2) \; C^{D  \to B^*}(0,t,T/2)}%
	     {C^{D  \to D^*}(0,t,T/2) \; C^{B  \to B^*}(0,t,T/2)}.
	\label{eq:RA1def}
\end{eqnarray}
Apart from renormalization factors, these ratios correspond to the
continuum ratios ${\cal R}_+$, ${\cal R}_1$, and ${\cal R}_{A_1}$.
In the window of time separations $t$ and ${T/2-t}$ for which the
lowest-lying states dominate, all convention-dependent normalization
factors cancel in the double ratios, and the ratios reduce to
\begin{eqnarray}
	        \rho_{V^{cb}}\sqrt{R_+}     & = & \sqrt{{\cal R}_+} =
				h_+(1),             \label{eq:rhoR=h+} \\
	        \rho_{V^{cb}}\sqrt{R_1}     & = & \sqrt{{\cal R}_1} =
				h_1(1),             \label{eq:rhoR=h1} \\
	\check{\rho}_{A^{cb}}\sqrt{R_{A_1}} & = & \sqrt{{\cal R}_{A_1}} =
				\check{h}_{A_1}(1), \label{eq:rhoR=hA1} 
\end{eqnarray}
where
$\check{\rho}_A^2=\rho_{A^{cb}}\rho_{A^{bc}}/\rho_{A^{cc}}\rho_{A^{bb}}$.
In particular, note that the axial current double ratio does not yield 
$h_{A_1}(1)$ directly, but instead $\check{h}_{A_1}(1)$, defined in
Eq.~(\ref{eq:contRA1}).
Once we have computed the left-hand sides of
Eqs.~(\ref{eq:rhoR=h+})--(\ref{eq:rhoR=hA1}) for several
combinations of the heavy quark masses, we can fit the mass
dependence to the form predicted by the heavy-quark expansions,
Eqs.~(\ref{eq:h+HQE})--(\ref{eq:checkhA1HQE}).

To summarize this section, let us review the steps needed to obtain the
physical form factor~${\cal F}_{B\to D^*}(1)$:
\begin{enumerate}
	\item compute the three-point correlation functions and thence the
	ratios~$R_+$, $R_1$, $R_{A_1}$;
	\item multiply $\sqrt{R}_+$ and $\sqrt{R}_1$ with $\rho_V/\eta_V$,
	and $\sqrt{R}_{A_1}$ with $\check{\rho}_A/\check{\eta}_A$, to
	obtain $h_+(1)/\eta_V$, $h_1(1)/\eta_V$,
	and~$\check{h}_{A_1}(1)/\check{\eta}_A$;
	\item fit $1-h/\eta$ [where $h/\eta$ is $h_+(1)/\eta_V$,
	$h_1(1)/\eta_V$, or~$\check{h}_{A_1}(1)/\check{\eta}_A$]
	to the heavy-quark mass dependence expected from
	Eqs.~(\ref{eq:h+HQE})--(\ref{eq:checkhA1HQE});
	\item use the resulting $\ell_V$, $\ell_A$, and $\ell_P$ 
	(and associated $1/m_Q^3$ terms) to reconstitute
	$h_{A_1}(1)={\cal F}_{B\to D^*}(1)$ via (the $1/m_Q^3$ version of)
	Eq.~(\ref{eq:hA1HQE}).
\end{enumerate}
As discussed above, with the lattice action, currents, and normalization
conditions chosen above, we obtain $h_{A_1}(1)$ with uncertainties of
order $\alpha_s(\bar{\Lambda}/2m_c)^2$ and~$\bar{\Lambda}^3/(2m_Q)^3$
from matching, although the fitting procedure also yields estimates
of three of the four $1/m_Q^3$ terms in~$h_{A_1}(1)$, as discussed
in Sec.~\ref{sec:results}.

\section{Lattice calculation}
\label{sec:Lattice_calculation}

This work uses three ensembles of lattice gauge field configurations,
which have been used in previous work on
heavy-light decay constants~\cite{Duncan:1995uq,El-Khadra:1998hq},
$B\to\pi l\nu$ and $D\to\pi l\nu$ semi-leptonic form
factors~\cite{Simone:1999ti},
light-quark masses~\cite{Gough:1997kw}, and
quarkonia~\cite{El-Khadra:1992vn}.
The quark propagators are the same as in Ref.~\cite{El-Khadra:1998hq}, 
but we now use 200 instead of 100 configurations on the finest lattice 
(with $\beta=6.1$).
The input parameters for these fields are in
Table~\ref{tab:sim_details}, together with some elementary output
parameters.
\begin{table}[tp]
\centering
\caption[tab:sim_details]{Input parameters to the numerical lattice
calculations, together with some elementary output parameters.
Error bars on the outputs refer to the last digit(s).}
\label{tab:sim_details} 
\begin{tabular}{lccc}
\multicolumn{4}{c}{Inputs} \\
\hline
$\beta=6/g_0^2$ &6.1 &5.9 &5.7 \\
Volume, $N_S^3\times N_T$ &$24^3\times48$ &$16^3\times32$ &$12^3\times24$\\
Configurations                  &200  &350 &300\\ 
$c_{\rm sw}$                    &1.46 &1.50 &1.57\\ 
$\kappa_h$, $m_0$~(GeV) &0.080, 7.90  &0.077, 6.03  &0.062, 6.16  \\ 
                        &0.090, 5.82  &0.088, 4.36  &0.089, 2.87  \\ 
                        &0.097, 4.62  &0.099, 3.06  &0.100, 2.03  \\
                        &0.100, 4.16  &0.110, 2.02  &0.110, 1.42  \\
                        &0.115, 2.21  &0.121, 1.16  &0.119, 0.96  \\
                        &0.122, 1.46  &0.126, 0.83  &0.125, 0.69  \\
                        &0.125, 1.16  &             &             \\
$\kappa_q$, $m_0$~(GeV) &0.1373, 0.092 &0.1385, 0.088 &0.1405, 0.093 \\ 
                        &0.1379, 0.039 &0.1388, 0.073 & \\ 
                        &              &0.1391, 0.057 & \\ 
$t$ range               & [9, 15]      & [6, 10]      & [4, 8] \\
\hline
\multicolumn{4}{c}{Elementary outputs} \\
\hline
$\kappa_{\text{crit}}$       & $0.13847^{+4}_{-2}$ & $0.14017^{+3}_{-1}$ &
	$0.14327^{+5}_{-2}$ \\
$a^{-1}_{\text{1P-1S}}$ (GeV) & $2.64^{+17}_{-13}$ & $1.81^{+7}_{-6}$    &
	$1.16^{+3}_{-3}$    \\ 
$a^{-1}_{f_\pi}$ (GeV)        & $2.40^{+10}_{-12}$ & $1.47^{+6}_{-6}$    &
	$0.89^{+2}_{-2}$    \\ 
$u_0$                         & 0.8816 & 0.8734 & 0.8608 \\
$\alpha_V(3.40/a)$            & 0.14533 & 0.15938 & 0.18265 \\
\end{tabular}
\end{table}

The quark propagators are computed from the Sheikholeslami-Wohlert (SW)
action~\cite{Sheikholeslami:1985ij}, which includes a dimension-five
interaction with coupling~$c_{\text{SW}}$, sometimes called the
``clover'' coupling.
For the light spectator quark we use customary normalization conditions
for massless quarks with the SW action, so $c_{\text{SW}}$ is adjusted
to reduce the leading lattice-spacing effect of Wilson fermions.
In practice, we adjust~$c_{\text{SW}}$ to the value $u_0^{-3}$
suggested by tadpole-improved, tree-level perturbation
theory~\cite{Lepage:1993xa}, and the so-called mean link $u_0$ is
calculated from the plaquette.
The leading light-quark cutoff effect is then of order
$\alpha_s\Lambda a$, multiplied by a numerical coefficient that is
known to be small.
For the heavy quarks we adjust~$c_{\text{SW}}$ to the same value, but,
as explained in Sec.~\ref{sec:bckgnd}, one should think of this
adjustment as tuning a coefficient in the HQET effective Lagrangian.

The hopping parameter~$\kappa$ is related to the bare quark mass.
For the heavy quarks, $\kappa_h$ is varied over a wide range
encompassing charm and bottom.
For the light spectator quark, the first row of $\kappa_q$ in
Table~\ref{tab:sim_details} corresponds to the strange quark.
To test the dependence of the form factors on the light quark mass,
we repeat the analysis for a few lighter spectator quarks.
Table~\ref{tab:sim_details} also lists the tadpole-improved bare quark
mass in~GeV,
\begin{equation}
	am_0 = \frac{1}{u_0} \left(\frac{1}{2\kappa}
		- \frac{1}{2\kappa_{\text{crit}}}\right),
	\label{eq:baremass}
\end{equation}
where the critical quark hopping parameter $\kappa_{\text{crit}}$
makes the pion massless.
Although this mass is just a bare mass, it shows that the heavy quarks
are heavy, and the light quarks~light.

The lattice spacing $a$ plays a minor role in our analysis, because
both the lattice perturbation theory and the fitting to the heavy-quark
mass dependence can be carried out in lattice units.
The physical scale enters only in adjusting the heavy-quark hopping
parameters to the physical mass spectra, and in studying the dependence
of~$h_{A_1}(1)$ on~$a$.
Table~\ref{tab:sim_details} contains two estimates of the lattice
spacing, from the spin-averaged 1P-1S splitting of charmonium,
$\Delta m_{\text{1P-1S}}$, and from the pion decay constant~$f_\pi$.

The renormalized strong coupling $\alpha_V(3.40/a)$ at scale~$3.40/a$
is determined as in Ref.~\cite{Lepage:1993xa}.
In Sec.~\ref{sec:PT} the coupling is run to $\alpha_V(q^*)$, where
$q^*$ is the optimal scale according to the Brodsky-Lepage-Mackenzie
(BLM) prescription~\cite{Brodsky:1983gc,Lepage:1993xa}.
Then $\alpha_V(q^*)$ is used to calculate the
short-distance coefficients $\rho_V/\eta_V$ and
$\check{\rho}_A/\check{\eta}_A$, which are introduced
in Eqs.~(\ref{eq:rhoR=h+})--(\ref{eq:rhoR=hA1}), as well as the
coefficient~$\eta_A$.

The right-hand side of Eq.~(\ref{three-ptB2D}) is the first term in a
series, with additional terms for each radial excitation [and similarly
for Eqs.~(\ref{three-ptB*2D*}) and~(\ref{three-ptB2D*})].
We reduce contamination from excited states in two ways.
First, we keep the three points of the three-point function well
separated in (Euclidean) time.
The initial-state meson creation operator is always at $t_i=0$
and the final-state meson annihilation operator at $t_f=N_T/2$.
We then vary the time $t_s$ of the current, to see when the lowest-lying
states dominate.
The second way to isolate the lowest-lying states is to choose creation
operators ${\cal O}^\dagger_{B^{(*)}}$ and annihilation operators
${\cal O}_{D^{(*)}}$ to provide a large overlap with the desired state.
This is done by smearing out the quark and anti-quark with 1S and 2S
Coulomb-gauge wave functions, as in Ref.~\cite{Duncan:1993eb}.

\begin{figure}[btp]%
\setlength{\PlotHeight}{0.27\textheight}\relax%
	\centering
	(a) \includegraphics[clip=true,height=\PlotHeight]{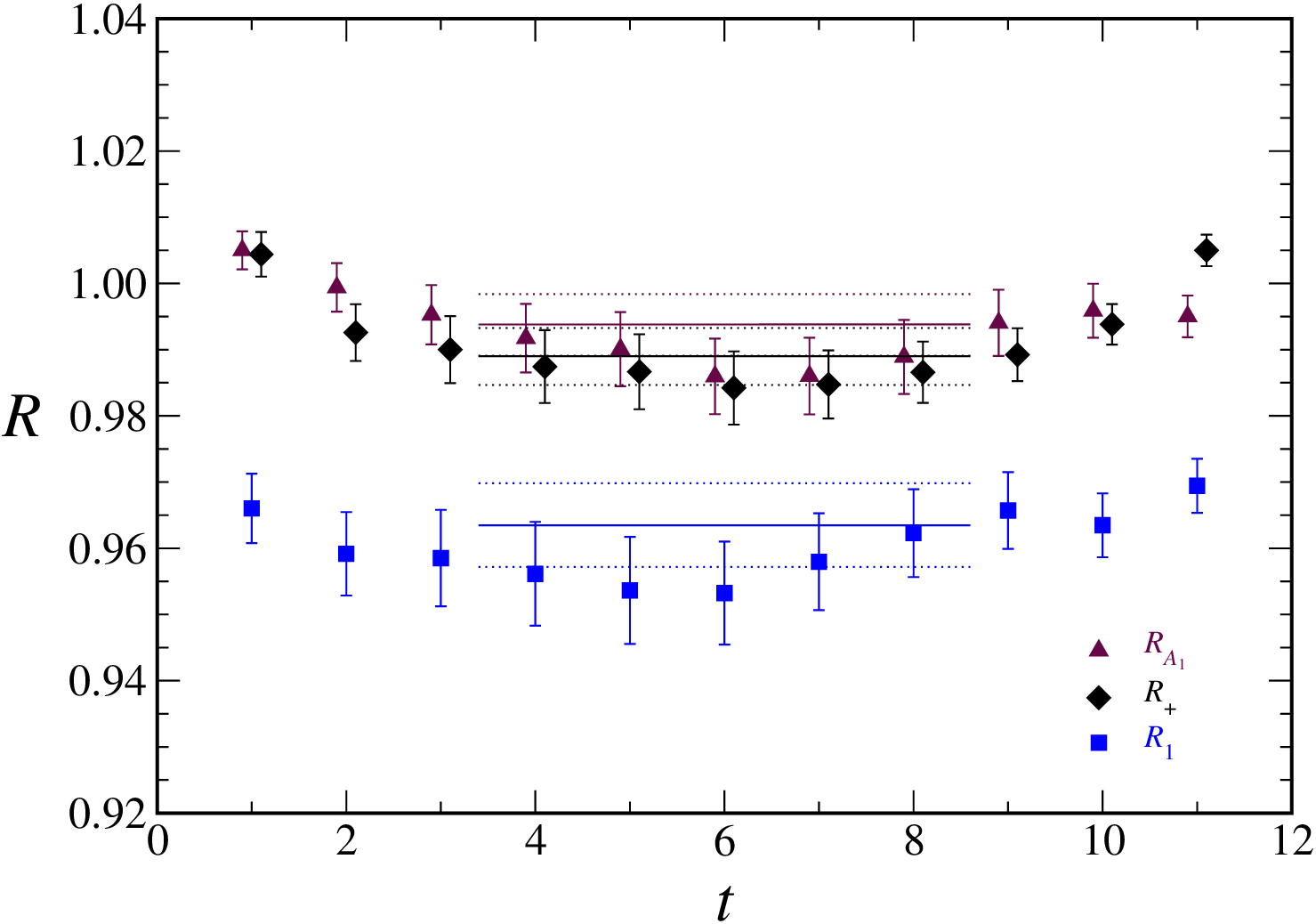}
	\vspace*{1.5em} \\
	(b) \includegraphics[clip=true,height=\PlotHeight]{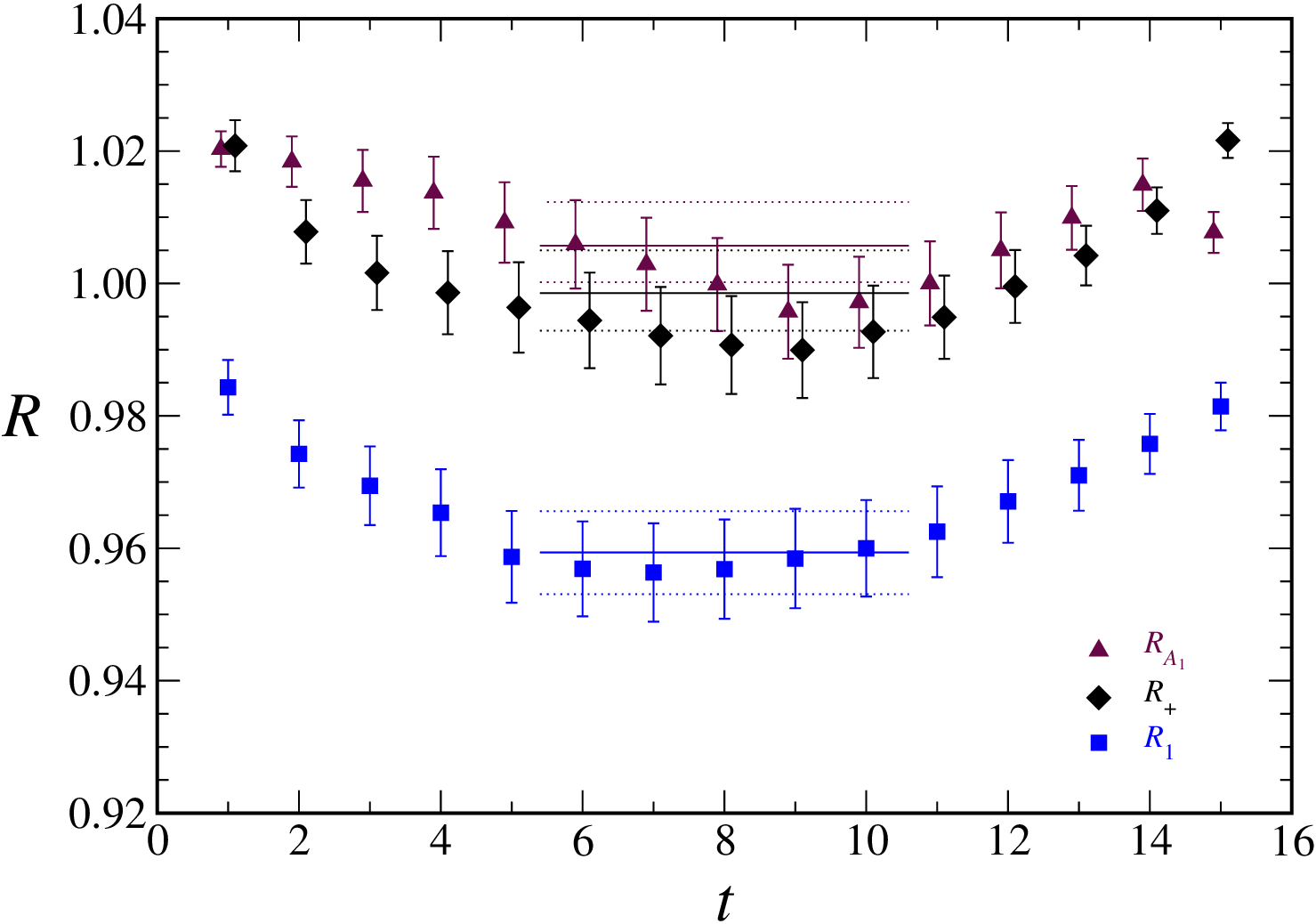}
	\vspace*{1.5em} \\
	(c) \includegraphics[clip=true,height=\PlotHeight]{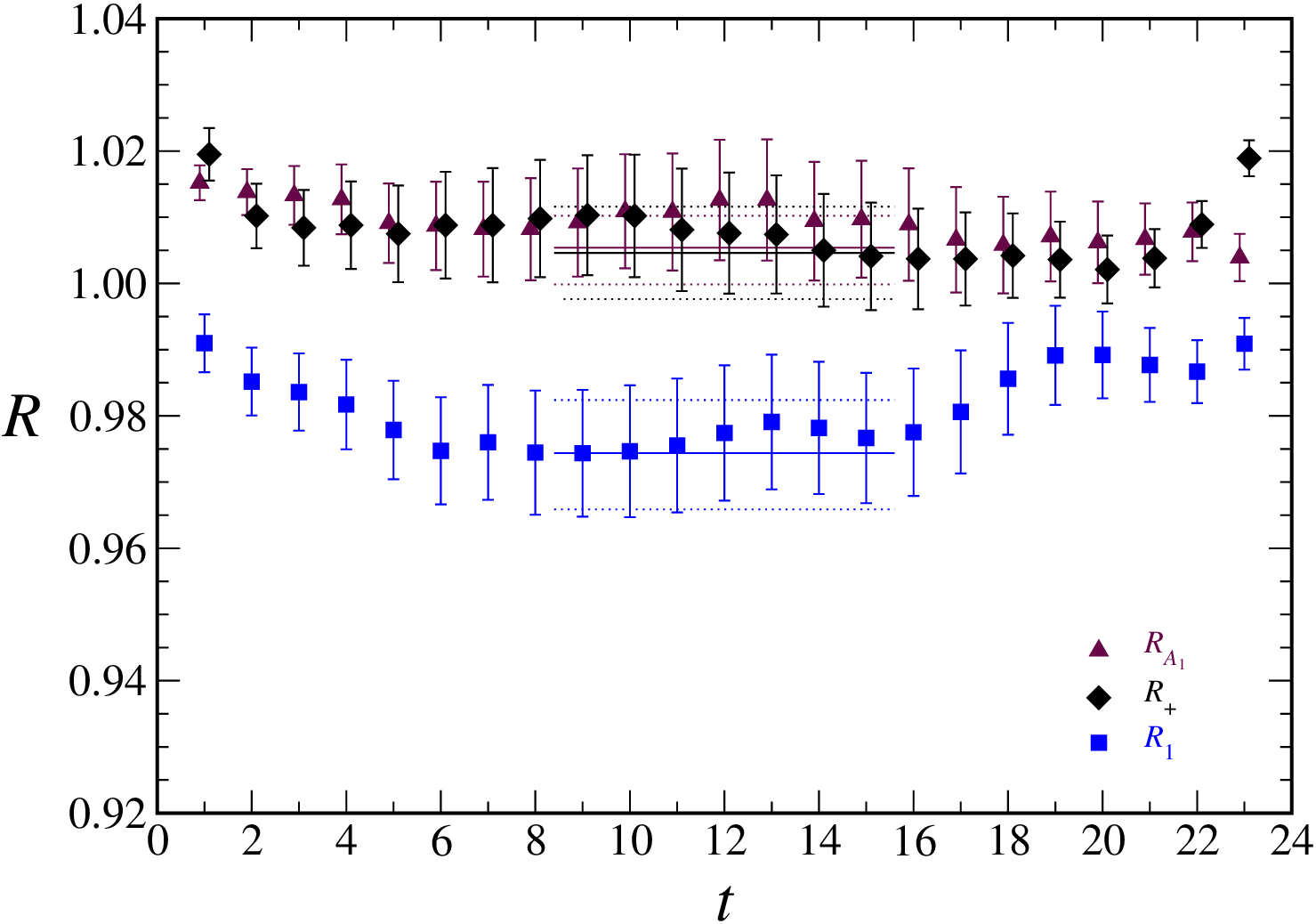}
	\vspace*{1em} \\
	\caption[fig:R]{Double ratios $R_{A_1}(t)$ (triangles),
		$R_+(t)$ (diamonds), and $R_1(t)$ (squares) at
		(a)~$\beta=5.7$, (b)~$\beta=5.9$, (c)~$\beta=6.1$.
		The heavy quark hopping parameters are
		(a)~$(\kappa_b,\kappa_c)=(0.062,0.100)$,
		(b)~$(\kappa_b,\kappa_c)=(0.088,0.121)$, and
		(c)~$(\kappa_b,\kappa_c)=(0.097,0.122)$.
		The light quark mass is close to the strange quark mass.
		The lines represent constant fits in the indicated ranges.}
	\label{fig:R}
\end{figure}

Figure~\ref{fig:R} shows the isolation of the ground state in the
ratios $R_+(t)$, $R_1(t)$, and $R_{A_1}(t)$.
In each of the three modes we find a long plateau.
We fit to a constant and obtain a precision at the percent level.
For each ensemble, we choose the same fit range for all mass
combinations listed in Table~\ref{tab:sim_details}.
In Fig.~\ref{fig:R} the resulting central values and error envelopes are
given by the solid and dotted lines, respectively.
Different fit ranges lead to slightly different, though consistent,
results; this variation is folded in with the statistical error.
Statistical errors, including the full correlation matrix in all fits,
are determined from 1000 bootstrap samples for each ensemble.
The bootstrap procedure is repeated with the same sequence for all quark
mass combinations, and in this way the fully correlated statistical
errors are propagated through all stages of the analysis.

Figure~\ref{fig:R} also demonstrates a clear distinction between the
$\bar{B}^*\to D^*$ and the other two modes.
Consequently, one can already see that $\ell_V$ is definitely
greater than~$\ell_P$ and~$\ell_A$, as expected from
Refs.~\cite{Neubert:1994vy,Shifman:1995jh,Bigi:1995ga}.
This is an important observation, because the largest $1/m_Q^2$
correction to~$h_{A_1}(1)$ is~$\ell_V/(2m_c)^2$.

\section{Perturbation theory}
\label{sec:PT}

\begin{figure}[btp]
	\centering
	\begin{picture}(200,120)(50,0)
	\thicklines
	\put(100,100){lattice}
	\put(112,95){\vector(0,-1){73}}
	\put(138,102){\vector(2,-1){67}}
	\put(200,54){HQET}
	\put(206,49){\vector(-2,-1){67}}
	\put(100,10){QCD}
	\put(90,54){$\rho$}
	\put(177,88){$\rho/\eta$}
	\put(177,23){$\eta$}
\end{picture}
	\caption[fig:matching]{Diagram illustrating how the matching factors
	$\rho$, $\eta$, and $\rho/\eta$ match lattice gauge theory and HQET
	to QCD, and to each other.}
	\label{fig:matching}
\end{figure}
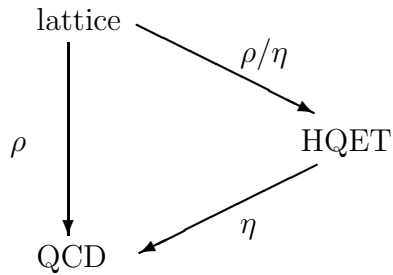

In this paper perturbation theory is needed to calculate the
short-distance coefficients $\rho_J$ ($J=V$, $A$) defined in
Eqs.~(\ref{eq:rhoVCC}) and~(\ref{eq:rhoACC}), and $\eta_J$
and $\check{\eta}_A$ appearing in Eqs.~(\ref{eq:hA1HQE})
and~(\ref{eq:h+HQE})--(\ref{eq:checkhA1HQE}).
The $\rho$ factors match lattice gauge theory to QCD, and the $\eta$
factors match HQET to QCD.
To fit the heavy-quark mass dependence of the lattice double ratios,
one must also match lattice gauge theory to HQET, and the corresponding
factors are simply $\rho_V/\eta_V$ and~$\check{\rho}_A/\check{\eta}_A$.
Figure~\ref{fig:matching} illustrates how these matching factors
connect lattice gauge theory and HQET to QCD, and to each other.

Lattice perturbation theory often yields a series that appears to
converge slowly.
The two main causes of the poor convergence have been
identified~\cite{Lepage:1993xa}: the bare gauge coupling is an
especially poor expansion parameter, and when tadpole diagrams occur
expansion coefficients are large.
These two problems can be avoided by using a renormalized coupling as
the expansion parameter and by using perturbation theory only for
quantities in which tadpole diagrams largely cancel.
Then lattice perturbation theory seems to converge as well as
perturbation theory in continuum~QCD.

To calculate the $\rho$ factors only the vertex function is needed.
By construction the self-energy contribution to wave-function
renormalization, in particular the tadpole diagrams, cancels completely.
Furthermore, even the vertex functions cancel partially, so the
expansion coefficients should be small, as verified explicitly at the
one-loop level~\cite{Kronfeld:1999tk}.
Indeed, as $m_Qa\to 0$, $\rho\to 1$, and as $m_Qa\to\infty$,
$\rho\to\eta$.
Thus, despite the fact that only the one-loop correction to $\rho_J$ is
available~\cite{Kronfeld:1999tk}, it seems likely that perturbation
theory can be expected to behave well, especially when measured against
other uncertainties in this calculation.

The other ingredient needed for an accurate perturbation series is a
suitable renormalized coupling.
We use the coupling~$\alpha_V$ defined through the (Fourier transform
of) the heavy quark potential, as suggested in Ref.~\cite{Lepage:1993xa}.
The scale~$q^*$ of the running coupling~$\alpha_V(q^*)$ is chosen
according to the BLM prescription~\cite{Brodsky:1983gc,Lepage:1993xa}:
\begin{equation}
	\log(q^*a)^2 = \frac{{}^*\zeta^{[1]}}{\zeta^{[1]}}.
	\label{eq:log(q*)}
\end{equation}
where $\zeta$ is $\rho_V/\eta_V$ or $\check{\rho}_A/\check{\eta}_A$
when fitting the mass dependence of the double ratios, or $\eta_A$
when reconstituting $h_{A_1}(1)$ with Eq.~(\ref{eq:hA1HQE}).
The numerator ${}^*\zeta^{[1]}$ in Eq.~(\ref{eq:log(q*)}) is obtained
from the Feynman integrand for $\zeta^{[1]}$ by replacing the gluon
propagator $D(k)$ by $\log(k^2a^2)D(k)$, where $k$ is the gluon's
momentum.
Such terms arise at the higher-loop level, so the BLM prescription sums
a class of higher-order corrections.
Since in the cases at hand the one-loop integrals are ultraviolet and
infrared finite, the only scales that can appear are $\sqrt{m_cm_b}$
and~$1/a$.
In general we find $q^*$ to be a few GeV; the only exceptions occur
when $(\rho_V/\eta_V)^{[1]}$ or $(\check{\rho}_A/\check{\eta}_A)^{[1]}$
are accidentally very small.

One of the advantages of the BLM prescription is that the scale
depends on the renormalization scheme, in such a way that the value of
the coupling itself does not depend on the scheme much.
The coupling in an arbitrary scheme~$S$ is related to the $V$~scheme
by
\begin{equation}
	\frac{(4\pi)^2}{g^2_S(q)} = \frac{(4\pi)^2}{g^2_V(q)} +
		\beta_0 b_S^{(1)} + b_S^{(0)} + O(g^2),
	\label{eq:bS}
\end{equation}
where for $n_f$ light quarks $\beta_0=11-2n_f/3$, 
and $b_S^{(0)}$ is independent of~$n_f$.
In many cases, the $\beta_0$ term dominates;
for example, for the $\overline{\rm MS}$ scheme,
$b_{\overline{\rm MS}}^{(1)}=-5/3$ and
$b_{\overline{\rm MS}}^{(0)}=-8$.
If one chooses $q_S^*=q^*e^{-b_S^{(1)}/2}$, then $g^2_S(q^*_S)$ differs
from $g^2_V(q^*)$ only by ``non-BLM'' terms of order~$g^4(\beta_0g^2)^{l-2}$,
$l\ge2$, which often are not very important.

In summary, we evaluate all short-distance coefficients with
\begin{equation}
	\zeta = 1 + \alpha_V(q^*) 4\pi\zeta^{[1]}
	\label{eq:PT}
\end{equation}
and the appropriate BLM scale~$q^*$.
To check for the possible size of non-BLM two-loop corrections (which
are unavailable for $\rho_J$), we also perform cross checks with
$\alpha_{\overline{\rm MS}}(q^*_{\overline{\rm MS}})$.
We obtain $\alpha_V(q^*)$ via two-loop running from~\cite{Lepage:1993xa}
\begin{equation}
	\alpha_V(3.40/a) = \frac{2\alpha_{1\times 1}}%
		{1 + \sqrt{1-4.74\alpha_{1\times 1}}} ,
\end{equation}
where $\alpha_{1\times 1}=-(3/\pi)\ln u_0$.
$u_0$ and $\alpha_V(3.40/a)$ are tabulated in
Table~\ref{tab:sim_details}.

Table~\ref{tab:R-and-rho} contains the values of $\rho_V/\eta_V$ and
$\check{\rho}_A/\check{\eta}_A$ appropriate to the heavy quark mass
combinations used in Sec.~\ref{sec:results}.
\begin{table}
	\centering
    \caption[tab:R-and-rho]{Double ratios, computed in the Monte
    Carlo, and (re)normalization factors, computed in perturbation
    theory to one-loop BLM order.}
    \label{tab:R-and-rho}
	\begin{tabular}{lllllll}
	$\beta$, $\kappa_q$
	       	& \multicolumn{1}{c}{$(\kappa_b,\kappa_c)$} 
			& \multicolumn{1}{c}{$\sqrt{R_+}$}
			& \multicolumn{1}{c}{$\sqrt{R_1}$}
			& \multicolumn{1}{c}{$\rho_V/\eta_V$}
			& \multicolumn{1}{c}{$\sqrt{R_{A_1}}$}
			& \multicolumn{1}{c}{$\check{\rho}_A/\check{\eta}_A$} \\
	\hline
	6.1	& (0.080, 0.115) & $1.0010^{+72}_{-75}$  
		& $0.9851^{+74}_{-77}$   & 1.0021 & $1.0024^{+68}_{-76}$   & 0.9940 \\
 0.1373 & (0.080, 0.122) & $1.0030^{+101}_{-102}$
		& $0.9742^{+106}_{-114}$ & 1.0008 & $1.0043^{+089}_{-106}$ & 0.9919 \\
		& (0.090, 0.100) & $1.0001^{+06}_{-06}$  
		& $0.9990^{+06}_{-06}$   & 1.0002 & $1.0002^{+06}_{-06}$   & 1.0000 \\
		& (0.090, 0.125) & $1.0050^{+70}_{-67}$  
		& $0.9757^{+81}_{-84}$   & 0.9978 & $1.0051^{+68}_{-68}$   & 0.9908 \\
		& (0.097, 0.115) & $1.0007^{+18}_{-18}$  
		& $0.9948^{+21}_{-21}$   & 1.0003 & $1.0012^{+16}_{-17}$   & 0.9985 \\
		& (0.097, 0.122) & $1.0023^{+35}_{-35}$  
		& $0.9871^{+41}_{-43}$   & 0.9991 & $1.0027^{+34}_{-34}$   & 0.9954 \\
		& (0.100, 0.125) & $1.0039^{+38}_{-36}$  
		& $0.9838^{+45}_{-47}$   & 0.9973 & $1.0034^{+36}_{-36}$   & 0.9933 \\
	\hline
	5.9	& (0.077, 0.110) & $0.9981^{+34}_{-28}$
		& $0.9872^{+33}_{-29}$   & 1.0030 & $1.0009^{+32}_{-27}$   & 1.0001 \\
 0.1385 & (0.077, 0.121) & $0.9971^{+58}_{-51}$
		& $0.9697^{+57}_{-54}$   & 1.0035 & $1.0030^{+57}_{-50}$   & 0.9770 \\
		& (0.077, 0.126) & $0.9984^{+69}_{-67}$
		& $0.9549^{+69}_{-71}$   & 1.0015 & $1.0054^{+70}_{-62}$   & 0.9868 \\
		& (0.088, 0.110) & $0.9993^{+15}_{-12}$
		& $0.9934^{+15}_{-13}$   & 1.0013 & $1.0007^{+14}_{-12}$   & 0.9999 \\
		& (0.088, 0.121) & $0.9993^{+32}_{-29}$
		& $0.9795^{+32}_{-32}$   & 1.0016 & $1.0028^{+33}_{-27}$   & 0.9944 \\
		& (0.088, 0.126) & $1.0011^{+46}_{-40}$
		& $0.9666^{+50}_{-47}$   & 0.9995 & $1.0053^{+44}_{-38}$   & 0.9903 \\
		& (0.099, 0.110) & $0.9999^{+04}_{-03}$
		& $0.9980^{+04}_{-03}$   & 1.0003 & $1.0003^{+04}_{-03}$   & 0.9990 \\
		& (0.099, 0.121) & $1.0003^{+16}_{-14}$
		& $0.9883^{+17}_{-16}$   & 1.0000 & $1.0019^{+15}_{-13}$   & 0.9969 \\
		& (0.099, 0.126) & $1.0022^{+27}_{-23}$
		& $0.9780^{+31}_{-28}$   & 0.9983 & $1.0041^{+25}_{-20}$   & 0.9983 \\
	\hline
	5.7	& (0.062, 0.089) & $0.9944^{+21}_{-26}$  
		& $0.9923^{+26}_{-28}$   & 1.0024 & $0.9975^{+23}_{-25}$   & 1.0010 \\
 0.1405 & (0.062, 0.100) & $0.9895^{+42}_{-43}$
		& $0.9845^{+50}_{-52}$   & 1.0050 & $0.9958^{+42}_{-48}$   & 1.0017 \\
		& (0.062, 0.125) & $0.9786^{+102}_{-118}$
		& $0.9339^{+122}_{-150}$ & 1.0114 & $0.9888^{+121}_{-118}$ & 1.0006 \\
		& (0.089, 0.100) & $0.9992^{+03}_{-03}$
		& $0.9984^{+04}_{-04}$   & 1.0005 & $0.9996^{+03}_{-03}$   & 1.0001 \\
		& (0.089, 0.110) & $0.9969^{+11}_{-10}$
		& $0.9929^{+15}_{-14}$   & 1.0018 & $0.9985^{+11}_{-11}$   & 1.0002 \\
		& (0.089, 0.119) & $0.9945^{+21}_{-22}$
		& $0.9816^{+32}_{-32}$   & 1.0035 & $0.9969^{+23}_{-24}$   & 1.0000 \\
		& (0.089, 0.125) & $0.9939^{+31}_{-34}$
		& $0.9673^{+50}_{-52}$   & 1.0041 & $0.9958^{+34}_{-37}$   & 1.0112 \\
		& (0.100, 0.125) & $0.9979^{+15}_{-18}$
		& $0.9793^{+29}_{-29}$   & 1.0022 & $0.9983^{+19}_{-21}$   & 0.9958 \\
		& (0.110, 0.119) & $0.9997^{+02}_{-02}$
		& $0.9972^{+04}_{-04}$   & 1.0004 & $0.9998^{+02}_{-03}$   & 0.9995 \\
    \end{tabular}
\end{table}

As expected, the perturbative corrections to these factors are small.
The lattice coefficients~$\rho_J^{[1]}$ and~${}^*\rho_J^{[1]}$ were
obtained in Ref.~\cite{Kronfeld:1999tk}.
The continuum coefficients are~\cite{Neubert:1995qt}
\begin{eqnarray}
	    \eta_V^{[1]}         & =  & C_F\, 3 f(m_b/m_c)/16\pi^2,
		\label{eq:etaV[1]}  \\
	{}^*\eta_V^{[1]}         & =  & C_F\, 9 f(m_b/m_c)/32\pi^2 +
		\eta_V^{[1]}\ln(m_bam_ca), \label{eq:*etaV[1]} \\
	    \check{\eta}_A^{[1]} & =  & C_F\, 3 f(m_b/m_c)/16\pi^2 ,
		\label{eq:check[1]} \\
	{}^*\check{\eta}_A^{[1]} & =  & C_F\, 5 f(m_b/m_c)/32\pi^2 +
		\check{\eta}_A^{[1]}\ln(m_bam_ca)
		, \label{eq:*check[1]}
\end{eqnarray}
where
\begin{equation}
	f(z) = \frac{z+1}{z-1}\ln z - 2.
	\label{eq:f}
\end{equation}
The important properties of $f(z)$ are $f(1)=0$, $f(1/z)=f(z)$.
From the matching procedure derived in Ref.~\cite{Kronfeld:1999tk} one
sees that the masses used in $f(m_b/m_c)$ should be the kinetic masses,
namely the mass appearing in the kinetic term in Eq.~(\ref{eq:LHQET}).

Two different schemes for defining the kinetic quark mass are used in
this paper, because they are simple to implement.
Both employ the formula~\cite{El-Khadra:1997mp}
\begin{equation}
	\frac{1}{am_2} = \frac{1}{e^{am_1}\sinh(am_1)} + \frac{1}{e^{am_1}},
	\label{eq:kinetic}
\end{equation}
which is the tree-level relation between the kinetic mass~$m_2$ and the
rest mass~$m_1$, for the SW action.
One choice is to use the tree-level value for the rest mass
$am_1=\log(1+am_0)$, with $am_0$ from Eq.~(\ref{eq:baremass}),
and we call the result the tree-level kinetic mass.
The other is to use the one-loop rest mass in Eq.~(\ref{eq:kinetic})
\cite{Mertens:1998wx}, and we call the result the quasi-one-loop
kinetic mass.
(The kinetic mass receives further radiative corrections, but they
are known to be small~\cite{Mertens:1998wx}.)
The second choice is essentially the (one-loop) perturbative pole mass.
Although the difference between these schemes is formally of the non-BLM
two-loop order, they could give slightly different results in practice.
Thus, using both and comparing gives us a handle on the terms omitted
from the perturbative series.

When reconstituting the physical form factor $h_{A_1}(1)$ with
Eq.~(\ref{eq:hA1HQE}), one needs a numerical value for the
short-distance coefficient~$\eta_A$.
Although it is known at the two-loop
level~\cite{Czarnecki:1996gu,Czarnecki:1997cf}, we use the one-loop, BLM
results, so that all perturbation theory is treated on the same footing.
Thus, we take~\cite{Neubert:1995qt}
\begin{eqnarray}
	    \eta_A^{[1]} & =  & C_F \left[ 3 f(m_b/m_c) - 2
			\right]/16\pi^2 , \label{eq:etaA[1]} \\
	{}^*\eta_A^{[1]} & =  & C_F \left[
		\case{5}{2} f(m_b/m_c) - 1 \right]/16\pi^2 +
		\eta_A^{[1]}\ln(m_bam_ca) . \label{eq:*etaA[1]}
\end{eqnarray}
For consistency, it is necessary to use the same definition of the
quark mass in~$\eta_A$ as in $\rho/\eta$.

If we take the quasi-one-loop kinetic masses, which are very close to
continuum pole masses, we find
$z=m_{2c}/m_{2b}=\{0.308, 0.296, 0.290\}$,
$q^*=\{2.94, 3.08, 3.12\}$~GeV, 
$\alpha_V(q^*)=\{0.205, 0.203, 0.208\}$ and, hence,
\begin{equation}
	\eta_A = \{ 0.9713, 0.9724, 0.9724 \}
	\label{eq:etaAa}
\end{equation}
for $\beta=\{5.7, 5.9, 6.1\}$, respectively.
%
%
On the other hand, if we take the tree-level kinetic masses,
we find $z=\{0.221, 0.230, 0.234\}$, $q^*=\{2.02, 2.14, 2.14\}$~GeV, 
$\alpha_V(q^*)=\{0.241, 0.238, 0.245\}$ and, hence,
\begin{equation}
	\eta_A = \{ 0.9769, 0.9758, 0.9746 \}
	\label{eq:etaAa-tree}
\end{equation}
for $\beta=\{5.7, 5.9, 6.1\}$, respectively.
%
%
Note that although the coupling is larger in this scheme (because the
quark masses and, hence, $q^*$ are smaller), the perturbative correction
is smaller, because the magnitude of the coefficient~$\eta_A^{[1]}$
decreases with~$z$.
As we shall see below, this scheme dependence in $\eta_A$ is largely
cancelled by the corresponding scheme dependence of the $1/m_Q^2$
corrections.

These values of~$\eta_A$ are slightly larger than the value
0.960~\cite{Czarnecki:1996gu,Czarnecki:1997cf},
which is widely adopted in the literature.
The origin of this difference is the value used for $\alpha_s$.
We extract $\alpha_s$ from lattice QCD, which, in
the quenched approximation, underestimates $\alpha_s$
slightly~\cite{El-Khadra:1992vn}.
Also, there is nothing special about the standard value.
It does not include uncertainties from the measured value of
$\alpha_s(M_Z)$ or from the $b$ and $c$ masses.
When our method is applied to full QCD, the double ratios, the gauge
coupling, and the quark masses all can be determined self-consistently.
In the meantime, we shall assign uncertainties from
omitting the non-BLM two-loop term,
adjusting the heavy quark masses, and
the quenching effect on~$\alpha_s$.

\section{Heavy quark mass dependence}
\label{sec:results}

In this section we fit the (suitably normalized) double ratios to the
form expected from the heavy quark expansion, yielding the quantities
$a^2\ell_V$, $a^2\ell_A$, and $a^2\ell_P$ (i.e., in lattice units).
We find that it is also necessary and beneficial to incorporate terms
of order $1/m_Q^3$ in the heavy quark expansion.
The last step is then to combine these results into the main goal, which
is $h_{A_1}(1)$.

Table~\ref{tab:R-and-rho} contains the results of our Monte Carlo
calculations of $\sqrt{R_+}$, $\sqrt{R_1}$, and $\sqrt{R_{A_1}}$,
in addition to the short-distance coefficients discussed in
Sec.~\ref{sec:PT}.
This information is combined to form
\begin{eqnarray}
	\frac{\rho_V\sqrt{R_+}}{\eta_V} & = & \frac{h_+}{\eta_V} ,
	\label{eq:rR/eta+} \\
	\frac{\rho_V\sqrt{R_1}}{\eta_V} & = & \frac{h_1}{\eta_V} ,
	\label{eq:rR/eta1} \\
	\frac{\check{\rho}_A\sqrt{R_{A_1}}}{\check{\eta}_A} & = &
		\frac{\check{h}_{A_1}}{\check{\eta}_A} ,
	\label{eq:rR/etaA1}
\end{eqnarray}
which we fit to the expected heavy-quark mass dependence.
For each ratio in Eqs.~(\ref{eq:rR/eta+})--(\ref{eq:rR/etaA1})
we try the fit
\begin{equation}
    \frac{\rho\sqrt{R}}{\eta} =
		1 -
		\case{1}{4} \Delta_2^2 \left( c^{(2)} +
		\case{1}{2} c^{(3)} \Sigma_2 \right),
  \label{eq:1/m_Q_expansion_general}
\end{equation}
where $c^{(2)}$ and $c^{(3)}$ are taken as free fit parameters, and 
\begin{eqnarray}
	\Delta_2 & = & \frac{1}{am_{2c}} - \frac{1}{am_{2b}},
	\label{eq:Delta} \\
	\Sigma_2 & = & \frac{1}{am_{2c}} + \frac{1}{am_{2b}}.
	\label{eq:Sigma}
\end{eqnarray}
In Eqs.~(\ref{eq:Delta}) and (\ref{eq:Sigma}), the subscript 2 indicates
that the kinetic mass~$m_2$ appears.
For the quadratic term we use $\Delta_2^2$, even though the masses
$m_2$, $m_{\cal B}$, and $m_3$ all appear in the heavy-quark expansion
to lattice QCD~\cite{Kronfeld:2000ck}, because $m_2=m_{\cal B}=m_3$
at our level of accuracy.
The rest mass~$m_1$ in Eq.~(\ref{eq:LHQET}) drops out
completely~\cite{Kronfeld:2000ck}.

The $1/m_Q^3$ term is introduced in Eq.~(\ref{eq:1/m_Q_expansion_general})
to describe the data over a wide range of~$1/m_Q$.
The particular form $\Delta^2\Sigma$ is the only one that is
invariant under the interchange symmetry $c\leftrightarrow b$ and
vanishes for $m_c=m_b$.
The $1/m_Q^3$ terms arise from many sources in~HQET.
Some of them, like triple insertions of the $1/m_Q$ terms in
${\cal L}_{\text{HQET}}$, are correctly normalized with the choice of
lattice action and currents made in Sec.~\ref{sec:Lattice_calculation}.
They lead to $\Delta_2^2\Sigma_2$, with (to our accuracy)
the kinetic mass everywhere.
Others, like an insertion of a $1/m_Q^2$ term combined with an insertion
of a $1/m_Q$ term, are not and would lead to $\Delta_2\Delta_X\Sigma_X$,
where $\Delta_X\Sigma_X$ amounts to the difference of short-distance
coefficients for the higher-dimension HQET operator~${\cal O}_X$.

The most important mismatches of $\Delta_X\Sigma_X$ are of
order~$\alpha_sam_{2c}$ and of order~$(am_{2c})^2$,
provided $am_{2c}<1$.
They are not necessarily small but, perhaps, small enough
to pin down the $1/m_Q^3$ corrections.
The $1/m_Q^3$ contributions are influenced mostly by the region with
large $\Sigma$, where $am_{2c}<0.6$.
Thus, the fit coefficients $c^{(3)}$ can be expected to give a
reasonable estimate of the desired $a^3\ell^{(3)}$.
Moreover, corrections of order $(\bar{\Lambda}/m_Q)^3$ are small to
begin with, so even a large relative uncertainty in them leads to a
small absolute uncertainty on~$h_{A_1}(1)$.

As mentioned in the Introduction, there are four $1/m_Q^3$ terms in the
heavy quark expansion of~$h_{A_1}(1)$.
If we write
\begin{equation}
	h_{A_1}(1) = \eta_A \left[1 +
		\delta_{1/m^2} + \delta_{1/m^3} \right],
	\label{eq:hA1HQE-3}
\end{equation}
then $\delta_{1/m^2}$ can be read off by comparing with
Eq.~(\ref{eq:hA1HQE}), and
\begin{equation}
	\delta_{1/m^3} = -
		\frac{\ell^{(3)}_V}{(2m_c)^3} +
		\frac{\ell^{(3)}_C}{(2m_c)^2(2m_b)} +
		\frac{\ell^{(3)}_B}{(2m_c)(2m_b)^2} -
		\frac{\ell^{(3)}_P}{(2m_b)^3} .
	\label{eq:delta3}
\end{equation}
As suggested by the notation, $\ell^{(3)}_V$ is related to $h_1(1)$,
and $\ell^{(3)}_P$ is related to $h_+(1)$.
Repeating the argument based on heavy-quark spin symmetry,
first for the~$b$, then for the~$c$, one sees that
$h_{A_1}(1)$ and $h_1(1)$ share the term $\ell^{(3)}_V/(2m_c)^3$,
and that
$h_{A_1}(1)$ and $h_+(1)$ share the term $\ell^{(3)}_P/(2m_b)^3$,
as given in Eq.~(\ref{eq:delta3}).
The other two terms in $\delta_{1/m^3}$ can be rewritten
\begin{equation}
	\frac{\ell^{(3)}_C}{(2m_c)^2(2m_b)} +
	\frac{\ell^{(3)}_B}{(2m_c)(2m_b)^2} =
		\frac{\ell^{(3)}_A}{(2m_c)(2m_b)}
		\left( \frac{1}{2m_c} + \frac{1}{2m_b} \right) +
		\frac{\ell^{(3)}_D}{(2m_c)(2m_b)} 
		\left( \frac{1}{2m_c} - \frac{1}{2m_b} \right) ,
\end{equation}
where
$\ell_A^{(3)}=\left[\ell^{(3)}_C+\ell^{(3)}_B\right]/2$ and
$\ell_D^{(3)}=\left[\ell^{(3)}_C-\ell^{(3)}_B\right]/2$.
Simple algebra shows that $\ell^{(3)}_A$ is indeed the coefficient
of the $\Delta^2\Sigma$ term in the heavy-quark expansion of the
ratio~$\check{h}_{A_1}(1)$.
Thus, to the extent that we can identify $c^{(3)}_{\{P,V,A\}}$ with
$a^3\ell^{(3)}_{\{P,V,A\}}$, we can reconstruct three of the four
$1/m_Q^3$ corrections to~$h_{A_1}(1)$.
Only $\ell_D^{(3)}$ eludes us.

\begin{figure}[btp]%
\setlength{\PlotHeight}{0.27\textheight}\relax%
	\centering
	(a) \includegraphics[clip=true,height=\PlotHeight]{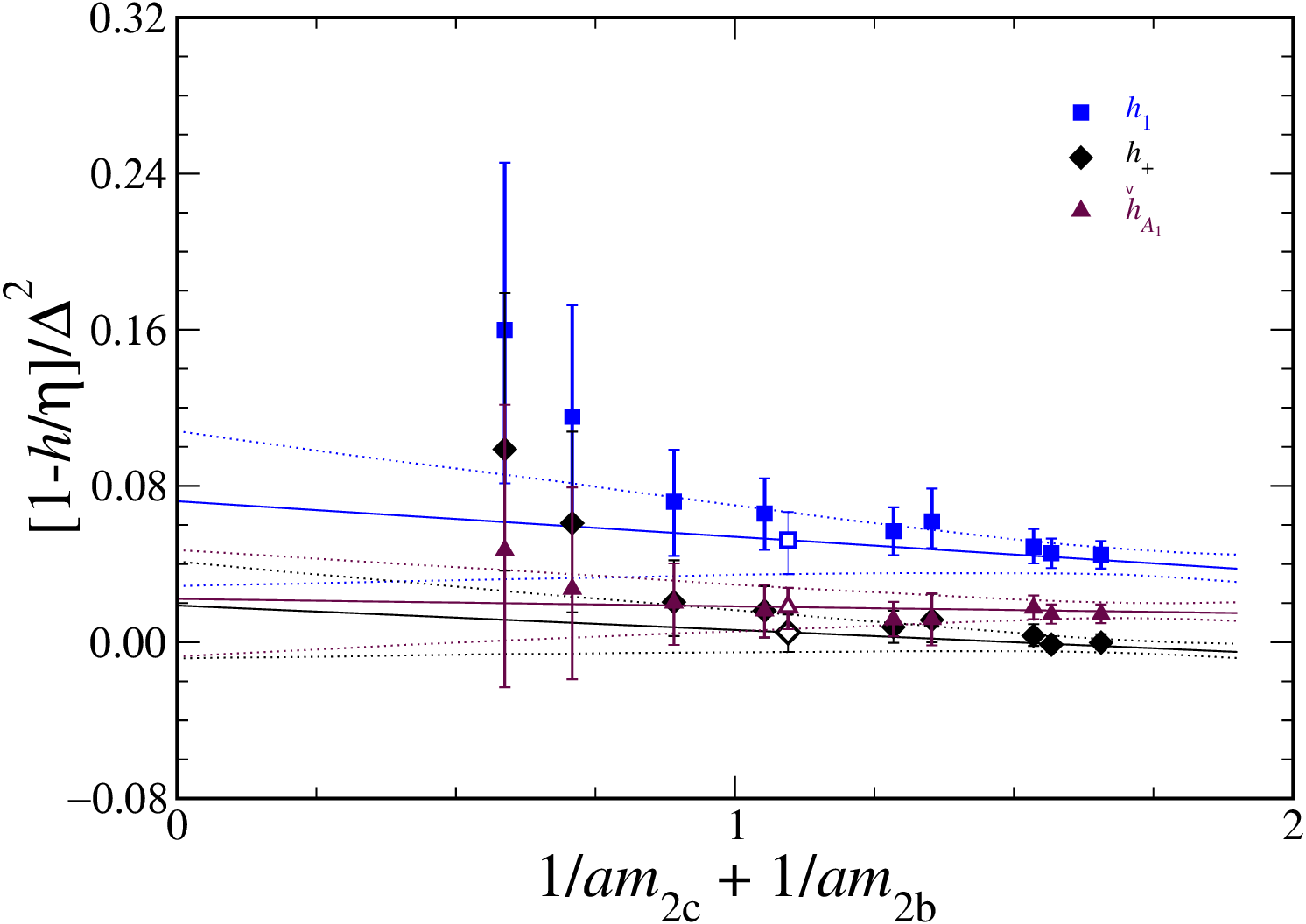}
	\vspace*{1.5em} \\
	(b) \includegraphics[clip=true,height=\PlotHeight]{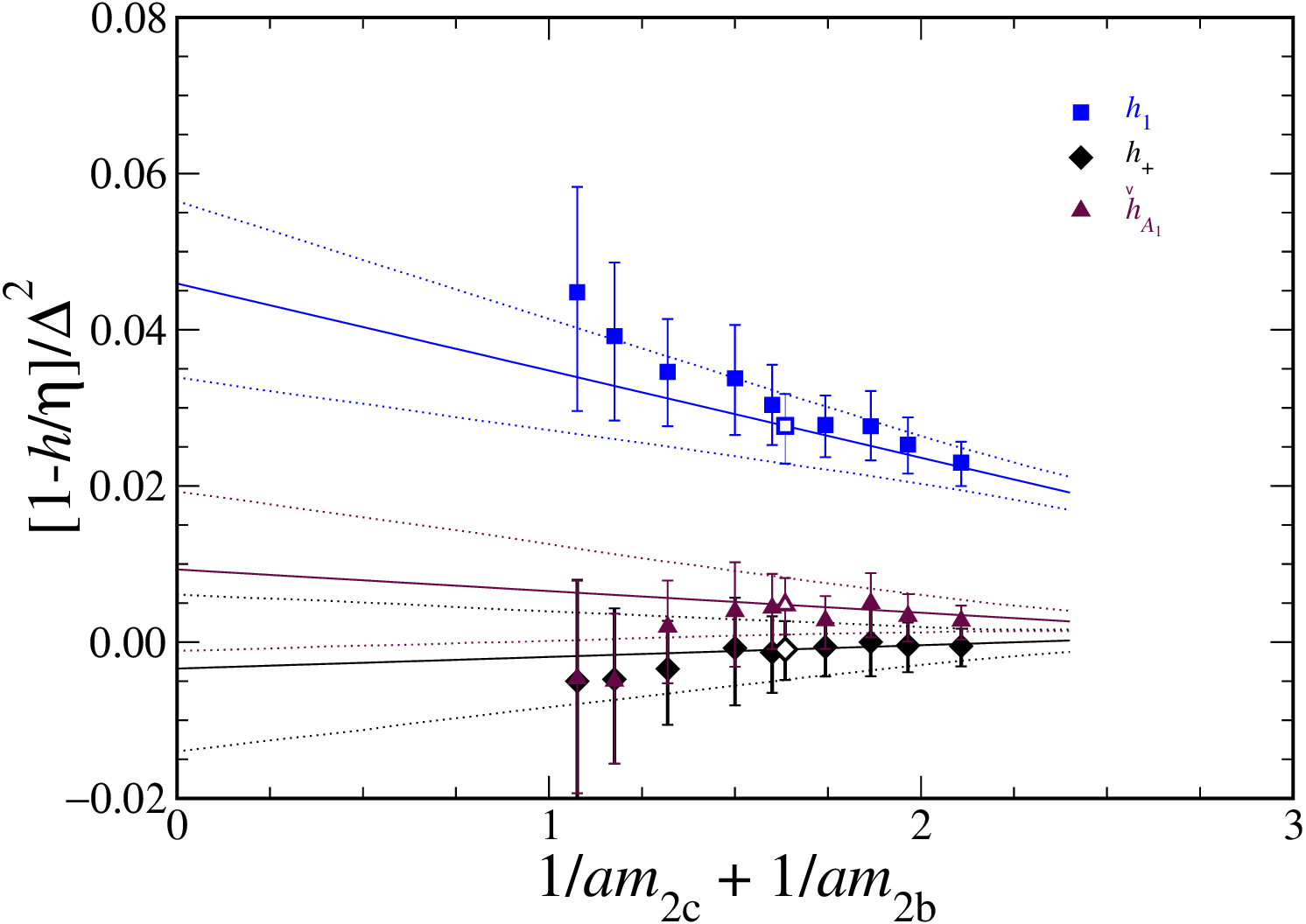}
	\vspace*{1.5em} \\
	(c) \includegraphics[clip=true,height=\PlotHeight]{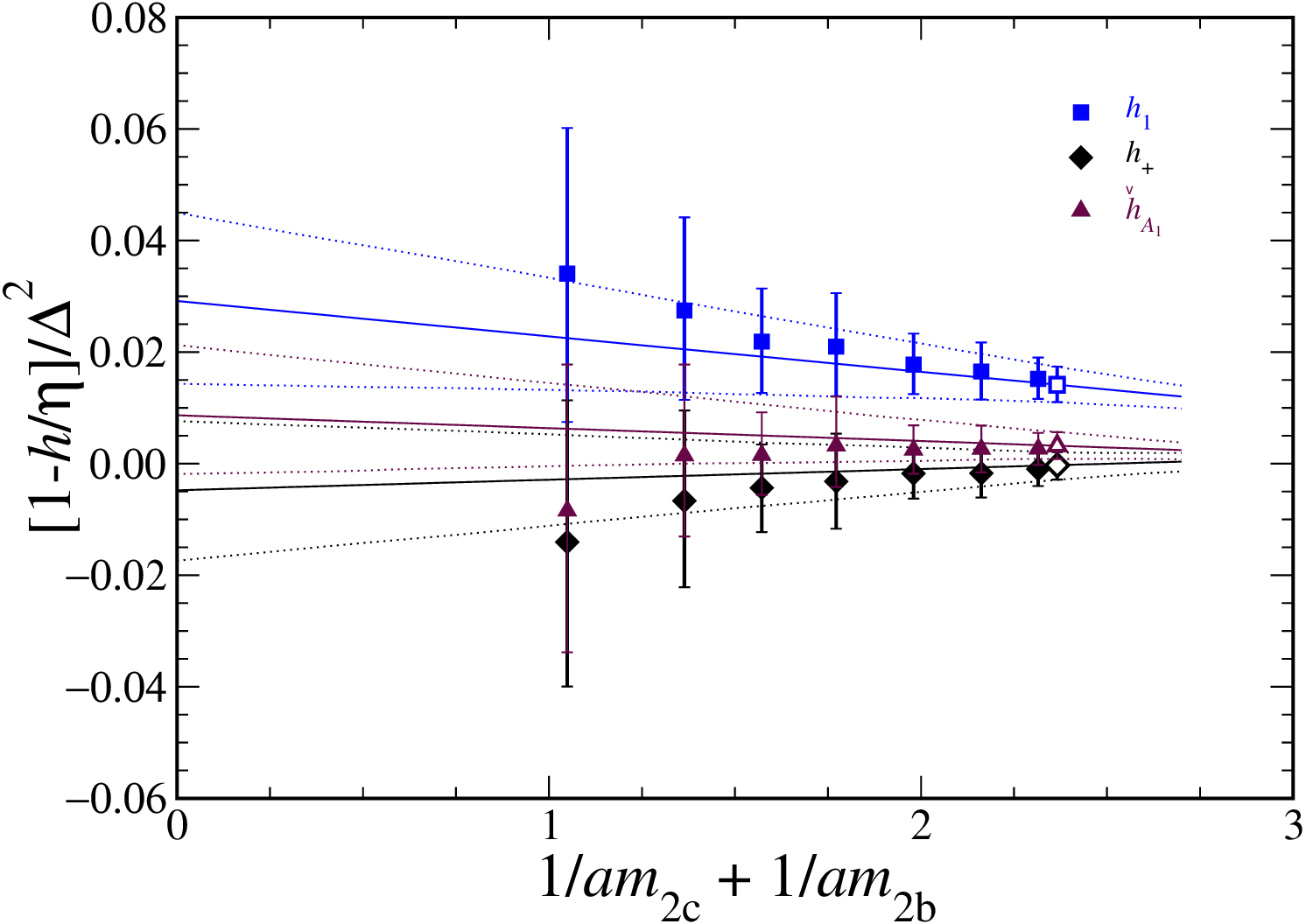}
	\vspace*{1em} \\
	\caption{
		$(1-h/\eta)/\Delta_2^2$ vs.\ $1/am_{2c}+1/am_{2b}$ when
		$h/\eta$ is $h_1(1)/\eta_V$ (squares),
		$h_+(1)/\eta_V$ (diamonds), and
		$\check{h}_{A_1}(1)/\check{\eta}_A$ (triangles) at
		(a)~$\beta=5.7$, (b)~$\beta=5.9$, (c)~$\beta=6.1$.
		Solid lines are best fits and dotted lines are error envelopes.}
    \label{fig:coeff}
\end{figure}

To show the quality of the fit to the mass dependence,
we plot in Fig.~\ref{fig:coeff} the quantity
\begin{equation}
	Q = \frac{1-\rho\sqrt{R}/\eta}{\Delta_2^2} =
		\case{1}{4} c^{(2)} + \case{1}{8} c^{(3)} \Sigma_2
	\label{eq:1/m_Q_dependence_reduced}
\end{equation}
vs.~$\Sigma_2=(1/am_{2c}+1/am_{2b})$, with the quasi-one-loop definition
of~$am_2$.
Linear behavior in $(1/am_{2c}+1/am_{2b})$ is observed for each form
factor, and we show the fit line in the figure.
Some curvature is noticeable for the heaviest masses in
Fig.~\ref{fig:coeff}(a), but the linear fit is still consistent within
statistical errors.
The growth of the statistical error toward the heavy-quark limit is
a property of the heavy-light meson in the Monte Carlo, and it is
unavoidable~\cite{Lepage:1992ui,Hashimoto:1994nd}.

The values of the fit
parameters~$c_{\{P,V,A\}}^{(2)}=a^2\ell_{\{P,V,A\}}$ and
$c_{\{P,V,A\}}^{(3)}$ are listed in Table~\ref{tab:coeff}.
\begin{table}[tbp]
	\centering
    \caption[tab:coeff]{Coefficients in the $1/m_Q$ expansion,
    	Eq.~(\ref{eq:1/m_Q_expansion_general}).} 
    \label{tab:coeff}
	\begin{tabular}{cr@{.}lr@{.}lr@{.}lr@{.}lr@{.}lr@{.}l}
		& \multicolumn{4}{c}{$h_+/\eta_V$} 
		& \multicolumn{4}{c}{$h_1/\eta_V$} 
		& \multicolumn{4}{c}{$\check{h}_{A_1}/\check{\eta}_A$} \\
	$\beta$	& \multicolumn{2}{c}{$c_P^{(2)}$} 
		& \multicolumn{2}{c}{$c_P^{(3)}$}
		& \multicolumn{2}{c}{$c_V^{(2)}$} 
		& \multicolumn{2}{c}{$c_V^{(3)}$}
		& \multicolumn{2}{c}{$c_A^{(2)}$} 
		& \multicolumn{2}{c}{$c_A^{(3)}$} \\
	\hline
	  6.1	\rule{0pt}{1.2em}
	  	& $-0$&$019^{+051}_{-050}$ &  $0$&$015^{+035}_{-035}$
		&  $0$&$117^{+063}_{-059}$ & $-0$&$051^{+045}_{-041}$
		&  $0$&$035^{+050}_{-042}$ & $-0$&$018^{+035}_{-029}$ \\
	  5.9	\rule{0pt}{1.2em}
	  	& $-0$&$014^{+042}_{-038}$ &  $0$&$012^{+033}_{-030}$
		&  $0$&$184^{+042}_{-048}$ & $-0$&$089^{+032}_{-036}$
		&  $0$&$037^{+040}_{-042}$ & $-0$&$022^{+032}_{-034}$ \\
	  5.7	\rule{0pt}{1.2em}
	  	&  $0$&$075^{+090}_{-108}$ & $-0$&$100^{+099}_{-123}$
		&  $0$&$289^{+144}_{-174}$ & $-0$&$145^{+158}_{-186}$
		&  $0$&$089^{+099}_{-118}$ & $-0$&$030^{+111}_{-137}$ \\
    \end{tabular}
\end{table}

In each case the extracted values of $c^{(2)}$ and $c^{(3)}$ are highly
correlated.
On the other hand, the combinations
\begin{eqnarray}
	a^2\ell_V^{\text{eff}} & = & c^{(2)}_V +
		\frac{c^{(3)}_V}{2am_{2c}}, \\
	a^2\ell_A^{\text{eff}} & = & c^{(2)}_A + \case{1}{2} c^{(3)}_A
		\left(\frac{1}{2am_{2c}}+\frac{1}{2am_{2b}}\right),
	\label{eq:mixed} \\
	a^2\ell_P^{\text{eff}} & = &
		c^{(2)}_P + \frac{c^{(3)}_P}{2am_{2b}},
\end{eqnarray}
are statistically more precise, because the correlated error cancels,
for the first two especially so.
These combinations appear directly in Eq.~(\ref{eq:hA1HQE-3}), provided
we can reliably identify $c^{(3)}_V$ with $a^3\ell^{(3)}_V$.
We argued above that this identification is not too bad, because the
coefficients $c^{(3)}$ should be influenced principally by smaller
masses.
As seen in Fig.~\ref{fig:coeff}, this predjudice is borne out, especially
when the correlated statistics are taken into account:
the best fits fit best for large $(1/am_{2c}+1/am_{2b})$.

The results presented in Fig.~\ref{fig:coeff} and Table~\ref{tab:coeff}
are all for the quasi-one-loop definition of~$am_2$.
One should keep in mind that the $\ell$s and $\ell^{(3)}$s have a
well-defined interpretation as matrix elements within HQET.
Their detailed definition depends on the renormalization
scheme of operators in HQET, as discussed, for example, in
Ref.~\cite{Kronfeld:2000gk}.
After reconstituting $h_{A_1}(1)$, however, the scheme chosen should
have only a minor, residual effect.
Repeating the fits with the tree-level definition of~$m_2a$
changes the fit coefficients significantly (as expected).
The change in $h_{A_1}(1)$ is, however, not great, and it is
of order~$\alpha_s/m_Q^2$, as expected.

To fix the physical values of $m_b$ and $m_c$ we compute the $B_s$
and $D_s$ spectra on the same ensembles of lattice gauge fields.
Combining these inputs with the second row of Table~\ref{tab:coeff}
($\beta=5.9$) (and omitting the $\ell_D^{(3)}$ contribution) we find
\begin{eqnarray}
	\delta_{1/m^n} & = & \delta_{1/m^2}+\delta_{1/m^3} \\
		& \simeq &
		- \frac{ \ell_V^{\text{eff}}}{(2m_c)^2}
		+ \frac{2\ell_A^{\text{eff}}}{2m_c\,2m_b} 
		- \frac{ \ell_P^{\text{eff}}}{(2m_b)^2} =
		-\left(0.0447^{+0.0078}_{-0.0070}\right),
  \label{eq:delta_1/m^2}
\end{eqnarray}
which is needed in Eq.~(\ref{eq:hA1HQE-3}).
The error quoted here is statistical only; systematic uncertainties
are considered in detail in the next section.
Equation~(\ref{eq:delta_1/m^2}) shows the power of our method:
even with 15\% statistical uncertainties on
$\delta_{1/m^n}=h_{A_1}/\eta_A-1$,
one can see that $h_{A_1}(1)$ itself can be very precise.

\section{Systematic errors}
\label{sec:errors}

The intermediate result in Eq.~(\ref{eq:delta_1/m^2}) is obtained at
one value of the lattice spacing, and with a spectator quark whose
mass is close to that of the strange quark.
In this section we consider the systematic uncertainty from varying $a$
and $m_q$, as well as those from other sources.
Table~\ref{tab:budget} summarizes the results of this analysis, giving
the absolute error on the main result, $h_{A_1}(1)$, and also fractional
error on~$1-h_{A_1}(1)$.
\begin{table}
	\centering
	\caption[tab:budget]{Budget of statistical and systematic
		uncertainties for $h_{A_1}(1)$ and $1-h_{A_1}(1)$.
		The row labeled ``total systematic'' does not include uncertainty
		from fitting, which is lumped with the statistical error.
		The statistical error is that after chiral extrapolation.}
	\label{tab:budget}
	\begin{tabular}{lllll}
		uncertainty & \multicolumn{2}{c}{$h_{A_1}$} &
		\multicolumn{2}{c}{$1-h_{A_1}$}  \\
		   &  &  & \multicolumn{2}{c}{(\%)} \\
		\hline
		statistics and fitting           &
			$+0.0238$ & $-0.0173$ & $+27$ & $-20$ \\
		adjusting $m_c$ and $m_b$        &
			$+0.0066$ & $-0.0068$ & $+~8$ & $-~8$ \\
		$\alpha_s^2$                     &
			\multicolumn{2}{c}{$\pm0.0082$} & \multicolumn{2}{c}{$\pm~9$} \\
		$\alpha_s(\bar{\Lambda}/2m_Q)^2$ &
			\multicolumn{2}{c}{$\pm0.0114$} & \multicolumn{2}{c}{$\pm13$} \\
		$(\bar{\Lambda})^3/(2m_Q)^3$     &
			\multicolumn{2}{c}{$\pm0.0017$} & \multicolumn{2}{c}{$\pm~2$}  \\
		$a$ dependence                   &
			$+0.0032$ & $-0.0141$ & $+~4$ & $-16$ \\
		chiral                           &
			$+0.0000$ & $-0.0163$ & $+~0$ & $-19$  \\
		quenching                        &
			$+0.0061$ & $-0.0143$ & $+~7$ & $-16$  \\
		\hline
		total systematic       &
			$+0.0171$ & $-0.0302$ & $+20$ & $-35$ \\
		total (stat $\oplus$ syst)       &
			$+0.0293$ & $-0.0349$ & $+34$ & $-40$ \\
	\end{tabular}
\end{table}

As noted above, the uncertainties should scale with~$1-h_{A_1}(1)$.

In the following subsections, we consider, in turn, the uncertainties
arising from fitting Ans\"atze, which incorporate contamination in
Eqs.~(\ref{three-ptB2D})--(\ref{three-ptB2D*}) of excited states
(Sec.~\ref{subsec:X});
heavy quark mass dependence (Sec.~\ref{subsec:heavy});
matching lattice gauge theory to HQET and QCD
(Sec.~\ref{subsec:matching});
lattice spacing dependence (Sec.~\ref{subsec:a});
light (spectator) quark mass dependence (Sec.~\ref{subsec:light}); and
the quenched approximation (Sec.~\ref{subsec:quench}).
In Table~\ref{tab:budget} the statistical uncertainty is added in
quadrature to that from fitting, as discussed in Sec.~\ref{subsec:X}.
As outlined in Sec.~\ref{sec:Lattice_calculation}, statistical
uncertainties are computed with the bootstrap method and full
covariance matrices.

\subsection{Fitting and excited states}
\label{subsec:X}

We define $\chi^2$ in our fits with the full covariance matrix.
For the plateau fits to $R(t)$
\begin{equation}
	\chi^2 = \sum_{t_1,t_2}
		\left[ R(t_1) - R_{\text{fit}} \right]
		\sigma^{-2}(t_1,t_2)
		\left[ R(t_2) - R_{\text{fit}} \right].
	\label{eq:chi2}
\end{equation}
Because the numerical data are so highly correlated, some components
of the (inverse) matrix $\sigma^{-2}(t_1,t_2)$ cannot be determined well.
These components are discarded, according to singular value
decomposition~(SVD), by eliminating eigenvectors of $\sigma^2$ whose
eigenvalue $\lambda<r_{\text{SVD}}\lambda_{\text{max}}$, with
$r_{\text{SVD}}$~small.
We find we have to set $r_{\text{SVD}}\sim 10^{-2}$ to remove the
noisy eigenvectors from $\chi^2$ in Eq.~(\ref{eq:chi2}).

A potential drawback of the double ratio technique is that an early
plateau could be induced.
We cope with this issue by trying many fit ranges for the time~$t_s$
of the current.
In general, fits to a constant have good $\chi^2$ and agree for
fit ranges within the plateaus clearly seen in Fig.~\ref{fig:R}.
For each ensemble of lattice gauge fields we choose a single range for
$t_s$ for all three ratios and all heavy quark mass combinations.
In each case, the range is chosen to give small statistical error on
$R_{\text{fit}}$, while maintaining a central value close to that from
short intervals centered on~$T/4$.

The expressions in Eqs.~(\ref{three-ptB2D})--(\ref{three-ptB2D*}),
relating three-point correlation functions to matrix elements,
suppress terms from radial excitations of the desired, lowest-lying
states.
Because of heavy-quark symmetry, corresponding excitations of the $D$
and $B$ systems have similar wave functions and mass splittings.
Consequently, their contribution to the double ratios largely cancels,
leaving a residue that is suppressed by $(\bar{\Lambda}/m_Q)^2$
as well as the exponential factor for large times.
Thus, the excited-state contamination in a double ratio scales as $R-1$,
rather than~$R$.

The fits of the heavy quark mass dependence are obtained by minimizing
\begin{equation}
	\chi^2 = \sum_{i,j} 
	  \left( Q_i - \case{1}{4}c^{(2)} - \case{1}{8}c^{(3)}{\Sigma_2}_i\right)
	  \sigma^{-2}_{ij}
	  \left( Q_j - \case{1}{4}c^{(2)} - \case{1}{8}c^{(3)}{\Sigma_2}_j\right),
\end{equation}
where $i$, $j$ label mass combinations.
Once again, not all components of $\sigma^{-2}$ are well determined.
The fits are stable with
$r_{\text{SVD}}=\{5\times 10^{-3}, 5\times 10^{-4}, 1\times 10^{-3}\}$
for $\beta=\{5.7, 5.9, 6.1\}$.

In summary, the fitting procedure to determine the double ratios $R_+$,
$R_1$, and $R_{A_1}$ depends on the fit range for $t_s$ and on the
cut~$r_{\text{SVD}}$ in the~SVD.
Similarly, the fit parameters of the heavy quark mass dependence,
$c^{(2)}$ and $c^{(3)}$, depend on an additional SVD~cut.
The central values quoted here are from the fit ranges given in
Table~\ref{tab:sim_details}, $r_{\text{SVD}}=10^{-2}$ for~$R(t)$,
and $r_{\text{SVD}}$ as given above for $c^{(2)}$ and~$c^{(3)}$.
We then repeat the analysis with larger and
smaller SVD~cuts and, for~$R(t)$, with other fit ranges.
The resulting variation in $h_{A_1}(1)$ is smaller than the statistical
error of the ``best fits''.
Since excited states contribute differently in each fitting Ansatz,
the uncertainty in fitting $R(t)$ incorporates the uncertainty due to
excited-state contamination.
For convenience in analyzing the other systematics, the fitting error
is added in quadrature to the statistical error.

\subsection{Heavy quark mass dependence}
\label{subsec:heavy}

The physical heavy quark masses enter when reconstituting $h_{A_1}$ with
Eq.~(\ref{eq:hA1HQE-3}).
We determine them by tuning the hopping parameters $\kappa_b$ and
$\kappa_c$ to reproduce the $B_s$ and $D_s$ spectra.
To do so, we must compute the meson kinetic masses, which are somewhat
noisy, and we must choose an observable to define the (inverse)
lattice spacing.
Thus, the tuned values of $\kappa_b$ and $\kappa_c$ have statistical
uncertainties, from both the meson masses and from~$a^{-1}$.

They also have systematic uncertainties.
For example, the inverse lattice spacing~$a^{-1}$ is not the same when
defined by the 1P-1S splitting of charmonium or by $f_\pi$, as noted
in Table~\ref{tab:sim_details}.
Similarly, $\kappa_b$ and $\kappa_c$ are not the same when quarkonium
spectra are used instead of heavy-light spectra, although for $\kappa_c$
this makes very little difference.
In the end, we are left with a range for $\kappa_b$ and $\kappa_c$
and, hence, the heavy quark masses used in Eq.~(\ref{eq:hA1HQE-3}).
This range leads to the error bar labeled ``adjusting $m_b$ and $m_c$''
in Table~\ref{tab:budget}.

\subsection{Matching}
\label{subsec:matching}

As discussed in Sec.~\ref{sec:bckgnd} our method for heavy quarks 
matches lattice gauge theory to QCD by normalizing the first few terms
in the heavy-quark expansion~\cite{El-Khadra:1997mp,Kronfeld:2000ck}.
This is necessary to keep heavy-quark discretization effects under
control, but the approximate nature of the (perturbative) matching
calculations leads to a series of uncertainties.
The three most important of these are listed in Table~\ref{tab:budget}.

The first is formally of order~$\alpha_s^2$.
It comes from omitting the non-BLM radiative corrections to the factors
$\rho_J$ and $\eta_J$ and from omitted loop corrections to the quark
masses and to~$\alpha_s$.
As discussed in Sec.~\ref{sec:PT}, $\rho_J$ comes from the cancellation
of (continuum and lattice) vertex functions.
Thus, by design, the coefficients of its perturbation series are
small---usually smaller than those in~$\eta_J$~\cite{Kronfeld:1999tk}.
With $\eta_A$ (and $\eta_V$) we can check explicitly how big the non-BLM
two-loop corrections are.
For example, the value of $h_{A_1}(1)$ is reduced by~0.0082 if we switch
to the $\overline{\rm MS}$ scheme and include the non-BLM two-loop part
of the~$\eta_J$.
Since the unknown two-loop corrections to the $\rho_J$ could
compensate, or even over-compensate, we take the two-loop uncertainty
to be~$\pm0.0082$.

The next matching uncertainty is formally of
order~$\alpha_s(\bar{\Lambda}/2m_c)^2$, from tuning the lattice action
and currents to HQET.
Setting $\alpha_s=0.2$, $\bar{\Lambda}=500$~MeV, and
$m_c=1.25$~GeV, one finds $\alpha_s(\bar{\Lambda}/2m_c)^2=0.008$.
Another way to estimate this effect is to compare the analysis with
tree-level heavy quark masses to the standard one with quasi-one-loop
masses.
The difference in $h_{A_1}(1)$ is in the same ballpark, at most +0.0114.
Since other schemes for the quark mass could lead to shifts in the other
direction, we take $\pm0.0114$ as the uncertainty from this source.

The last matching uncertainty is of order $(\bar{\Lambda}/m_Q)^3$, from
the omission~of
\begin{equation}
	\frac{\ell^{(3)}_D}{(2m_c)(2m_b)} 
	\left( \frac{1}{2m_c} - \frac{1}{2m_b} \right) \sim 0.0017,
	\label{eq:17}
\end{equation}
assuming $\ell_D^{(3)}=\bar{\Lambda}^3$, $m_b=4$~GeV, and the same
values as above.
With same choices made above, we estimate that
$\ell^{(3)}_A[1/(2m_c)+1/(2m_b)]/(2m_c2m_b)$ and
$\ell^{(3)}_P/(2m_b)^3$ 
should be around 0.0033, and 0.0002, respectively.
In Table~\ref{tab:compare} we show the actual effect of the included
$1/m_Q^3$ corrections.
The scatter of the different analyses bears out the latter estimates,
lending credence to Eq.~(\ref{eq:17}).
\begin{table}
	\centering
	\caption[tab:compare]{Scheme dependence of $h_{A_1}(1)$.
		For each value of~$\beta$, the columns compare the scheme
		with tree-level and quasi-one-loop kinetic masses in $\eta_J$
		and in the mass dependence.
		The rows compare the effect of the $1/m^3$ contributions;
		here $\Sigma=1/m_c+1/m_b$ refers to the correction in
		Eq.~(\ref{eq:mixed}).
		Each row includes the corrections from all preceding~rows.}
	\label{tab:compare}
	\begin{tabular}{lllllll}
		\multicolumn{1}{c}{$1/m^n$} &
		\multicolumn{2}{c}{$\beta=6.1$} &
		\multicolumn{2}{c}{$\beta=5.9$} &
		\multicolumn{2}{c}{$\beta=5.7$} \\
		& \multicolumn{1}{c}{tree} & \multicolumn{1}{c}{quasi} &
		  \multicolumn{1}{c}{tree} & \multicolumn{1}{c}{quasi} &
		  \multicolumn{1}{c}{tree} & \multicolumn{1}{c}{quasi} \\
		\hline
		$1/m_Q^2$         \rule{0pt}{1.2em} &
			$0.8755^{+0.0343}_{-0.0372}$ & 
			$0.8948^{+0.0416}_{-0.0430}$ & 
			$0.9078^{+0.0113}_{-0.0097}$ & 
			$0.9103^{+0.0140}_{-0.0130}$ & 
			$0.9365^{+0.0173}_{-0.0141}$ & 
			$0.9303^{+0.0275}_{-0.0234}$ \\ 
		$1/m_c^3$         \rule{0pt}{1.2em} &
			$0.9331^{+0.0150}_{-0.0123}$ & 
			$0.9329^{+0.0205}_{-0.0167}$ & 
			$0.9362^{+0.0056}_{-0.0051}$ & 
			$0.9321^{+0.0082}_{-0.0072}$ & 
			$0.9549^{+0.0099}_{-0.0086}$ & 
			$0.9406^{+0.0162}_{-0.0151}$ \\
		$1/m_b^3$         \rule{0pt}{1.2em} &
			$0.9332^{+0.0150}_{-0.0124}$ & 
			$0.9326^{+0.0206}_{-0.0166}$ & 
			$0.9363^{+0.0056}_{-0.0051}$ & 
			$0.9320^{+0.0082}_{-0.0073}$ & 
			$0.9551^{+0.0099}_{-0.0086}$ & 
			$0.9409^{+0.0163}_{-0.0151}$ \\
		$\Sigma/(m_cm_b)$ \rule{0pt}{1.2em} &
			$0.9275^{+0.0126}_{-0.0114}$ & 
			$0.9274^{+0.0163}_{-0.0148}$ & 
			$0.9338^{+0.0057}_{-0.0052}$ & 
			$0.9300^{+0.0076}_{-0.0068}$ & 
			$0.9503^{+0.0097}_{-0.0079}$ & 
			$0.9400^{+0.0152}_{-0.0135}$ \\
	\end{tabular}
\end{table}

Uncertainties in the included $1/m_Q^3$ terms are smaller than
Eq.~(\ref{eq:17}), because many of them are obtained correctly, and
the mismatch in the others is small.

\subsection{Lattice spacing dependence}
\label{subsec:a}

\begin{figure}[btp]%
\setlength{\PlotHeight}{0.3\textheight}\relax%
	\centering
	\includegraphics[clip=true,height=\PlotHeight]{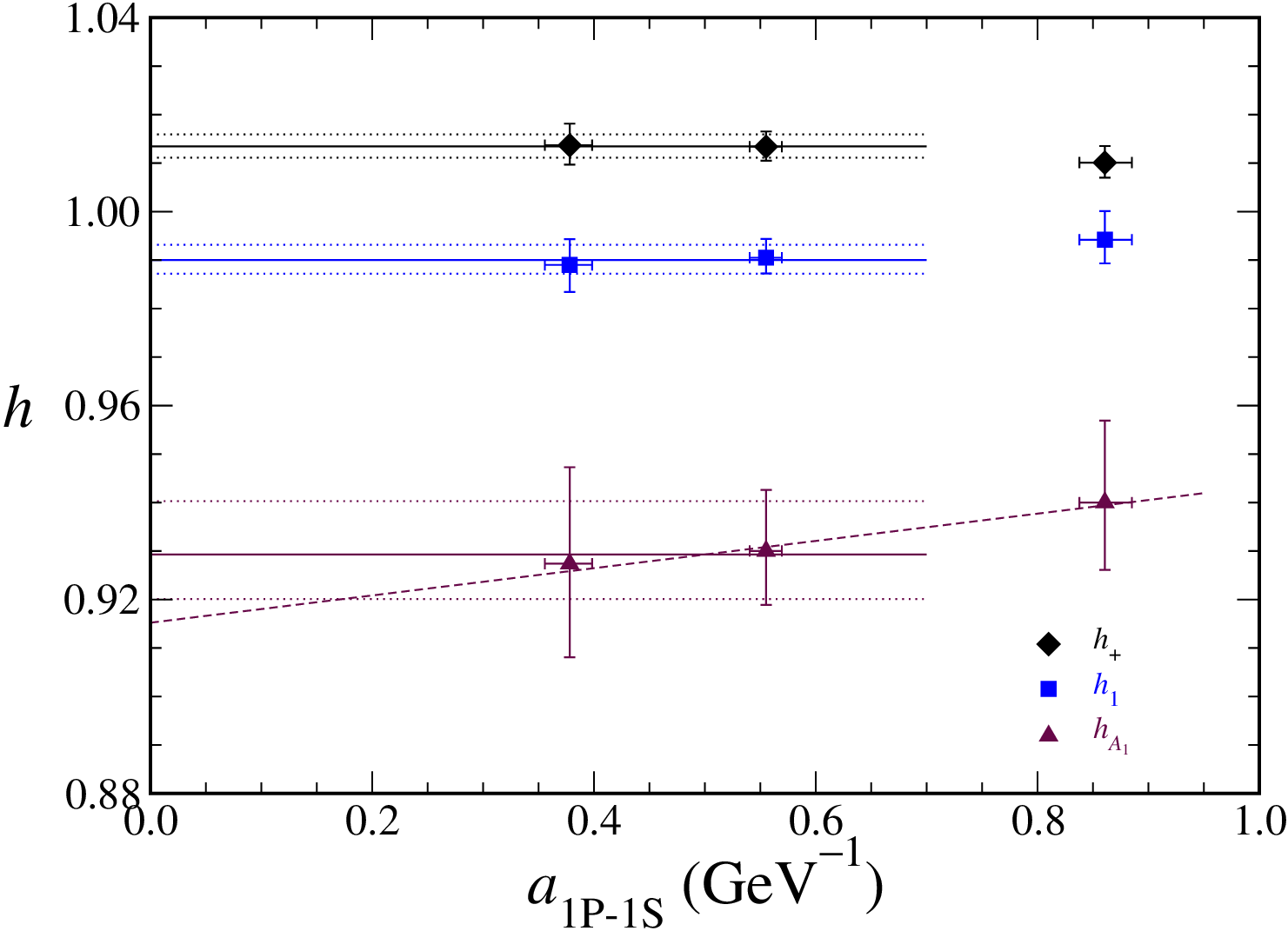}
	\caption[fig:h_vs_a]{Lattice spacing dependence of
		$h_{A_1}(1)$ (triangles), $h_+(1)$ (diamonds), and
		$h_1(1)$ (squares).
		The light quark mass is close to the strange quark mass.
		The solid (dotted) lines represent best fits (error envelopes).}
	\label{fig:h_vs_a}
\end{figure}

The lattice calculation of $h_{A_1}$ has lattice artifacts from heavy
quarks, light quarks, and gluons.
For the heavy quarks, discretization effects and heavy-quark effects are
inevitably intertwined~\cite{El-Khadra:1997mp,Kronfeld:2000ck},
and are mostly part and parcel of the matching uncertainties considered
above.
The light quarks suffer from discretization effects of
order~$\alpha_s\Lambda a$ and $(\Lambda a)^2$;
the gluons of order~$(\Lambda a)^2$.
That being said, we can test for the magnitude of discretization
effects, by comparing the analysis of Sec.~\ref{sec:results} for three
lattice spacings.
The results are plotted against~$a$ in Fig.~\ref{fig:h_vs_a},
which also contains results for $h_+(1)$ and $h_1(1)$.
The last two are much closer to~1 and their statistical uncertainties
are correspondingly smaller.
This underscores, once again, that the uncertainties scale as $1-h$.

The result for $h_{A_1}(1)$ with the available $1/m_Q^3$ contributions
(solid triangles) is consistent with a constant.
We take as our central value the average from the two finer lattices,
because for them the (heavy-quark) discretization effects are smaller.
This is
\begin{equation}
	h_{A_1}(1) = 0.9293^{+0.0110}_{-0.0092}
	\label{eq:hA1avg}
\end{equation}
where the error is the statistical error on the average, with
the error from fitting added in quadrature.
In Fig.~\ref{fig:h_vs_a} the solid and dotted lines indicate this
average and error band.

The third point, at $a=0.84~\text{GeV}$ (from $\beta=5.7$), has the
greatest uncertainty from \emph{heavy} quark discretization effects,
so it is excluded from the central value.
Instead we use it to estimate discretization uncertainties.
If one assumes that discretization effects from the light spectator
quark and gluons are negligible, then it would be appropriate to
average all three.
This average is slightly higher, and we take this increase as the
upward systematic error bar.
If, on the other hand, one assumes that the light spectator quark's
discretization effects are responsible for the somewhat larger value of
$h_{A_1}(1)$ on the coarsest lattice, then it would be appropriate to
extrapolate linearly in~$a$.
The dashed line in Fig.~\ref{fig:h_vs_a} shows this extrapolation.
The extrapolated value is significantly lower, and we take this decrease
as the downward systematic error bar.
The error bar resulting from these two estimates is very
asymmetric:~$^{+0.0032}_{-0.0141}$.

\subsection{Chiral extrapolation}
\label{subsec:light}

\begin{figure}[btp]%
\setlength{\PlotHeight}{0.3\textheight}\relax%
	\centering
	\includegraphics[clip=true,height=\PlotHeight]{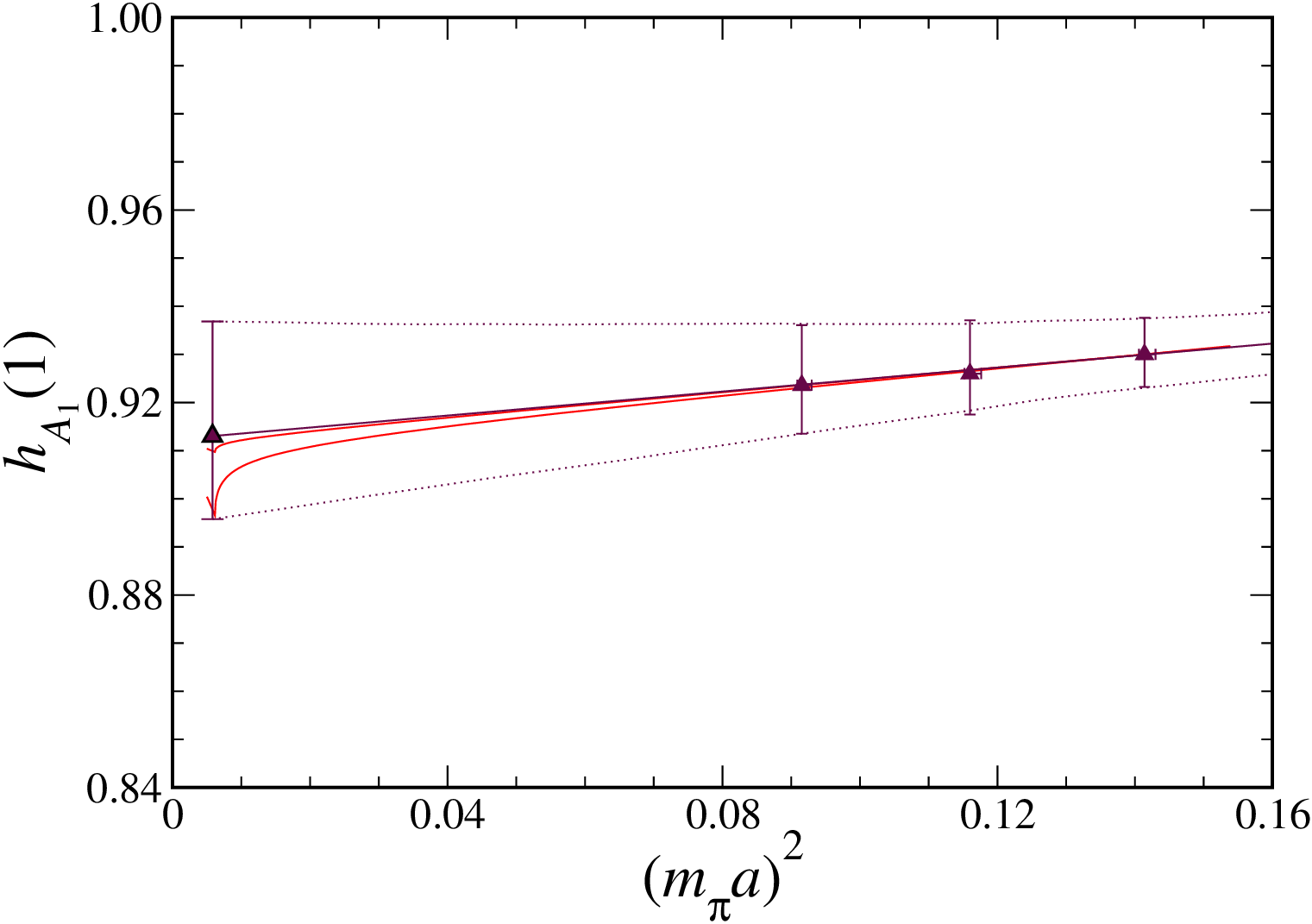}
	\caption[fig:chiral]{Dependence of $h_{A_1}(1)$
		at $\beta=5.9$ on the mass of the light spectator quark.
		Here $m_\pi^2$ is the mass of the pseudoscalar meson
		consisting of two ``light'' quarks.
		The solid (dotted) lines represent the best linear fit
		(error envelope).
		The lower (upper) curves with a cusp add to the linear
		behavior the contribution in Eq.~(\ref{eq:chiral}), taking
		$g_{D^*D\pi}=0.60$ ($g_{D^*D\pi}=0.27$).}
	\label{fig:chiral}
\end{figure}

The calculations discussed so far have a spectator quark whose mass
is near that of the strange quark.
Figure~\ref{fig:chiral} shows how $h_{A_1}(1)$ changes for lighter
spectator quarks, on the lattice with $\beta=5.9$, for which we have
three values of the light quark mass.
$h_{A_1}(1)$ is plotted against $m_\pi^2$ (in lattice units),
which is a physical measure of the light quark mass.
Since the statistical errors in Fig.~\ref{fig:chiral} are highly
correlated, the downward trend in~$h_{A_1}(1)$ is significant.
The same trend is seen for $\beta=6.1$.
Extrapolating linearly in $m_\pi^2$ to the physical pion,
reduces the result in Eq.~(\ref{eq:hA1avg}) to
\begin{equation}
	h_{A_1}(1) = 0.9130^{+0.0238}_{-0.0173}
	\label{eq:hA1Xtrap}
\end{equation}
and increases the statistical error.
This value, using the average of the $\beta=5.9$ and $6.1$ lattices
and the chiral extrapolation from $\beta=5.9$, gives the central value
in Eq.~(\ref{eq:result}).

In the chiral expansion, the terms responsible for the linear behavior
are formally of order $\bar{\Lambda}^2 m_\pi^2/(2m_c\,4\pi f_\pi)^2$.
Terms of order $\bar{\Lambda}^4/(2m_c\,4\pi f_\pi)^2$ are larger for
the physical pion mass, but are comparable for our artificially
large pion masses.
Randall and Wise~\cite{Randall:1993qg} have computed the $m_\pi^0$~effect
at one loop in chiral perturbation theory.
They find
\begin{equation}
	\frac{\ell_V(m_{\pi})}{(2m_c)^2} =
		\frac{\ell_V(m_{\eta_s})}{(2m_c)^2} +
	        \frac{g_{D^*D\pi}^2}{2}
			\left(\frac{\Delta^{(c)}}{4\pi f_\pi}\right)^2
			\left[\ln \frac{m_{\eta_s}^2}{m_\pi^2} + f(-x_\pi) -
			f(-x_{\eta_s}) \right]
	\label{eq:chiral}
\end{equation}
where $m_{\eta_s}^2=2m_K^2$ is the mass of the pseudoscalar meson with
two strange quarks, $g_{D^*D\pi}$ is the $D^*$-$D$-$\pi$ coupling,
$\Delta^{(c)}=m_{D^*}-m_D=142$~MeV is the $D^*$-$D$ mass splitting,
and $x_a=\Delta^{(c)}/m_a$ ($a=\pi$, $\eta_s$).
For~$g_{D^*D\pi}$ we consider the range 0.27--0.60, which encompasses
estimates based on fits to experimental data
($g_{D^*D\pi}=0.27^{+0.06}_{-0.03}$~\cite{Stewart:1998ke}),
quark models ($g_{D^*D\pi}\approx0.38$~\cite{Casalbuoni:1997pg}),
quenched lattice QCD ($g_{D^*D\pi}=0.30\pm0.16$~\cite{Aoki:2001rd}
or $g_{B^*B\pi}=0.42\pm0.09$~\cite{deDivitiis:2000mr}),
and the recent measurement of the $D^*$ width
($g_{D^*D\pi}=0.59\pm0.07$~\cite{Anastassov:2001cw}).

The chiral loop function $f(x)$ has rather different behavior,
depending on~$x$.
At $x=-1$, which turns out to be the physical region
($m_\pi\approx\Delta^{(c)}$), there is a cusp, and the value of $f$
becomes large: $f(-1)\approx11$ whereas $f(-x_{\eta_s})=f(-0.2)=0.53$.
To illustrate this behavior, we have shown in Fig.~\ref{fig:chiral}
the sum of the second term in Eq.~(\ref{eq:chiral}) and the linear fit.
In the region where we have data, the term from Eq.~(\ref{eq:chiral})
hardly varies, but near the physical limit, it bends the curve down.
With the quoted range for $g_{D^*D\pi}$, the decrease in~$h_{A_1}(1)$
amounts to~0.0033--0.0163, coming mostly in the region where
$m_\pi\approx\Delta^{(c)}$, as shown in Fig.~\ref{fig:chiral}.
In an unquenched calculation, one would add this contribution
to~$h_{A_1}(1)$.
Because $g_{D^*D\pi}$ remains uncertain and because we are using the
quenched approximation, we take it as an additional systematic
uncertainty of~$^{+0.0000}_{-0.0163}$.
This effect and the amplification of the statistical error together
make the chiral extrapolation the largest source of uncertainty.

\subsection{Quenching}
\label{subsec:quench}

An important limitation of our numerical value for $h_{A_1}(1)$ is that
the gauge fields were generated in the quenched approximation.
The quenched approximation omits the back-reaction of light quark loops
on the gluons, and compensates the omission with a shift in the bare
couplings.
Two obvious consequences of quenching are that the coupling $\alpha_s$
runs incorrectly, and that pion loops [as in Eq.~(\ref{eq:chiral})]
are not correctly generated.

Let us consider first the effect on the running coupling.
The values for $\eta_A$ in Sec.~\ref{sec:PT} are obtained with the
quenched coupling.
If $\alpha_s$ is corrected for quenching, it is
larger~\cite{El-Khadra:1992vn}, and the short-distance coefficients are
changed by $-0.0050$ for $\eta_A$ and $+0.0032$ for~$\eta_V$.
These changes both reduce $h_{A_1}(1)$.

For the pion-loop contribution we can look to comparisons of quenched
and unquenched calculations of other matrix elements.
Studies of the decay constants $f_B$ and $f_D$ show discrepancies on the
order of 10\% between quenched and (partly) unquenched
QCD~\cite{Bernard:1998xi,AliKhan:2001eg}.
A~form factor, which is the overlap of two wave functions, is presumably
less sensitive to quenching than a decay constant, which is a wave
function at the origin.
So, one should not expect the quenching error here to be more than~10\%.
Even in the quenched approximation all three double ratios tend to
unity in the heavy-quark symmetry limit.
Thus, the quenching error, like all others, scales with $R-1$,
rather than~$R$.
We therefore apply the estimate of 10\% to the long-distance part,
$\delta_{1/m^n}$, to obtain an error bar of~$\pm0.0061$.

We estimate the total quenching uncertainty to be the sum of these two
effects, or~$^{+0.0061}_{-0.0143}$.

\subsection{Summary}
\label{subsec:summary}

Combining Eq.~(\ref{eq:hA1Xtrap}) with the error budget in
Table~\ref{tab:budget}, we obtain
\begin{equation}
	h_{A_1}(1) = 0.9130^{+0.0238}_{-0.0173}
		{}^{+0.0156}_{-0.0157}
		{}^{+0.0032}_{-0.0141}
		{}^{+0.0000}_{-0.0163}
		{}^{+0.0061}_{-0.0143},
	\label{eq:final}
\end{equation}
where the error bars are from
statistics and fitting,
adjusting the heavy quark masses and matching,
lattice spacing dependence,
light quark mass dependence,
and the quenched approximation.
(The uncertainties on the second through fifth rows of
Table~\ref{tab:budget} are added in quadrature.)
Adding all systematics in quadrature, we obtain
\begin{equation}
	h_{A_1}(1) = {\cal F}_{B\to D^*}(1) =
		0.9130^{+0.0238}_{-0.0173}{}^{+0.0171}_{-0.0302}.
	\label{eq:quad}
\end{equation}
Although we have considered all sources of systematic uncertainty, it is
not possible to disentangle them completely.
For example, the lattice spacing dependence is not completely separated
from the HQET matching uncertainties, and the quenched approximation
affects the chiral behavior, the adjustment of $m_c$ and $m_b$,
and, through $\alpha_s$, the matching coefficients.

\section{Comparison with other methods}
\label{sec:comparison}

In this section we compare our method, based on lattice gauge theory,
with others existing in the literature.
To do so, it is convenient to refer to Eq.~(\ref{eq:hA1HQE}) and discuss
how the short- and long-distance contributions are evaluated.

One approach, sometimes advertised as ``model-independent'',
is to estimate the $\ell$s with the non-relativistic quark
model~\cite{Falk:1993wt,Neubert:1994vy}.
The more recent estimate~\cite{Neubert:1994vy} takes $\delta_{1/m^2}$ to
be $-0.055\pm0.025$ by covering a range of ``all reasonable choices''.
Combining it with the two-loop calculation~\cite{Czarnecki:1996gu}
of $\eta_A$, one obtains 
\begin{equation}
  {\cal F}_{B\to D^*}(1)=0.907 \pm 0.007 \pm 0.025 \pm 0.017,
  \label{eq:quark}
\end{equation}
where the quoted uncertainties~\cite{Neubert:1994vy,Czarnecki:1996gu}
are from perturbation theory,
errors in the quark model estimate of the $1/m_Q^2$ terms,
and the omission of $1/m_Q^3$ terms.
Uncertainties from $\alpha_s$ and the quark masses are not included.
A fair criticism of this approach is that it does not pay close
attention to scheme dependence of the long- and short-distance
contributions.
The standard ($\mu$-independent) result for $\eta_A$ corresponds to
renormalizing the operator insertions of~HQET in the
$\overline{\rm MS}$~scheme.
The quark model estimates, on the other hand, are presumably in some
other scheme, so there is a possibility to over- or undercount the
contribution at the interface of long and short distances.

Another approach is based on a zero-recoil sum
rule~\cite{Shifman:1995jh,Bigi:1997fj}.
These authors prefer to introduce a concrete separation scale~$\mu$.
In this scheme $\eta_A$ and the $\ell$s depend explicitly on~$\mu$.
The $\mu$-dependent two-loop part of $\eta_A$ is
known~\cite{Czarnecki:1998wy}.
A~recent estimate of the zero-recoil form factor
is~\cite{Uraltsev:2000qw}
\begin{equation}
  {\cal F}_{B\to D^*}(1)=0.89 \pm 0.015 \pm 0.025 \pm 0.015
  	\pm 0.025,
  \label{eq:rule}
\end{equation}
where the quoted uncertainties are from
the unknown value of the kinetic energy~$\mu^2_\pi(\mu)$,
higher excitations with $D^*$ quantum numbers and energy $E<m_{D^*}+\mu$,
perturbation theory,
and the omission of $1/m_Q^3$ terms.
We note that both $\mu^2_\pi$ and the excitation contribution should, in
this scheme, cancel the $(\mu/m_Q)^2$ part of $\eta_A(\mu)$.
Since there is no model-independent method to calculate the excitation
contribution (except unquenched lattice QCD), it is not clear how to
implement this cancellation.

\begin{figure}[!b]%
\setlength{\PlotHeight}{3.0in}\relax%
	\centering
	\includegraphics[clip=true,height=\PlotHeight]{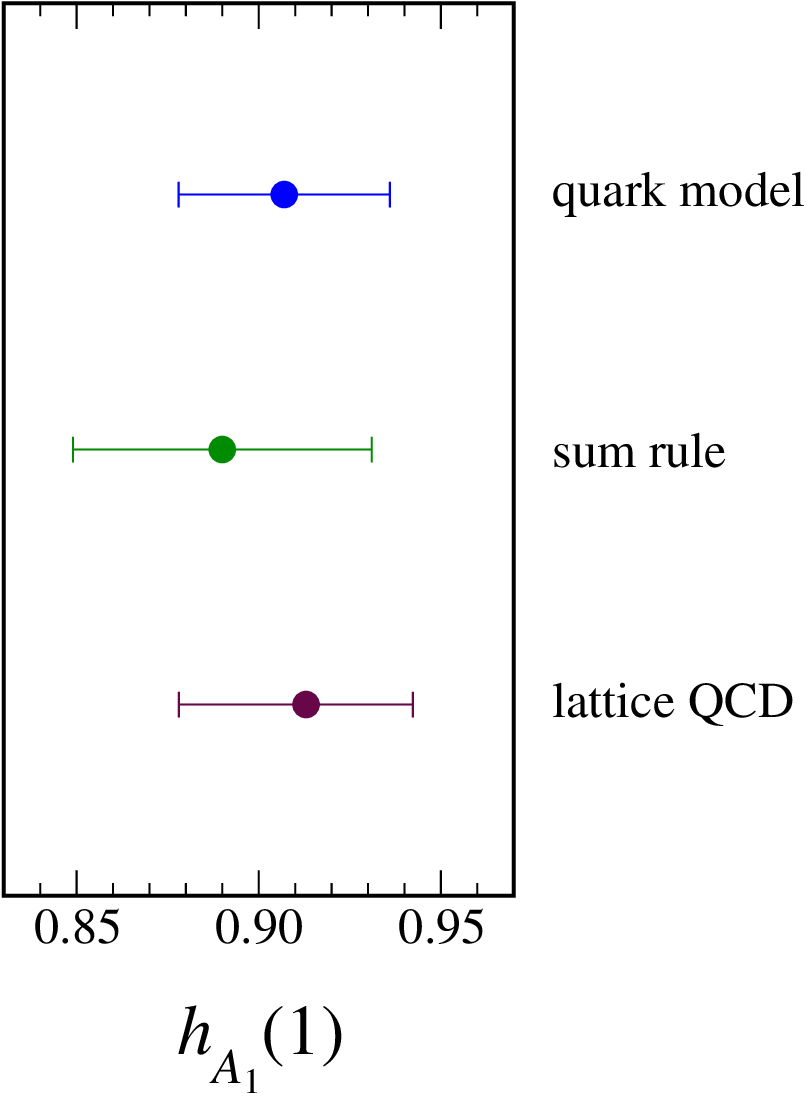}
	\caption[fig:compare]{Comparison of determinations of
		$h_{A_1}(1)={\cal F}_{B\to D^*}(1)$ with non-perturbative
		input from the non-relativistic quark
		model~\cite{Neubert:1994vy,Czarnecki:1996gu},
		a zero-recoil sum rule~\cite{Uraltsev:2000qw},
		and quenched lattice QCD.}
	\label{fig:compare}
\end{figure}

As shown in Fig.~\ref{fig:compare}, our result Eq.~(\ref{eq:result})
agrees with the previous results, within errors, and the quoted errors
are of comparable size.
Our result includes an estimate of three of the four $1/m_Q^3$
contributions.
All three are subject to a QED correction of +0.007~\cite{BaBar}.
An important feature of our method is that, even in the quenched
approximation, we are able to separate long- and short-distance
contributions self-consistently.
Indeed, we have repeated the calculation with two different schemes for
the heavy quark masses, and the results are the same.
Furthermore, it is clear that moving terms of order $\mu^2/m_Q^2$
between the long- and short-distance parts will cancel out in our
method, as long as it is done consistently.
Finally, with future unquenched calculations in lattice QCD, our
method allows for a systematic reduction in the theoretical error
on~$|V_{cb}|$.

\section{Conclusions}
\label{sec:conclusion}

We have developed a method to calculate the zero recoil form
factor of $\bar{B}\to D^*l\nu$ decay.
We introduce three double ratios in which the bulk of statistical
and systematic errors cancels, thus enabling a precise calculation of
${\cal F}_{B\to D^*}(1)$.
By matching lattice gauge theory to HQET, we are able to separate
long-distance from short-distance contributions.
Then the coefficients in the $1/m_Q$ expansion are obtained by
fitting the numerical data.
In this way we obtain the (leading) $1/m_Q^2$ corrections and three of
the four $1/m_Q^3$ corrections.
A~similar approach has already been taken for
$B\to Dl\nu$~\cite{Hashimoto:2000yp}.

Our result in the quenched approximation,
${\cal F}_{B\to D^*}(1)=0.913^{+0.024}_{-0.017}{}^{+0.017}_{-0.030}$,
is consistent with results based on other ways of treating
non-perturbative~QCD.
By using the quenched approximation we are able to gain control over
all other uncertainties.
Note, however, that the second error bar incorporates (among others)
our estimate of the uncertainty from quenching.
Furthermore, despite the shortcomings of the quenched approximation,
it is not less rigorous than competing determinations of
${\cal F}_{B\to D^*}(1)$, which use either non-relativistic quark models
or a subjective estimate of the ``excitation contribution''.
With recent measurements of $|V_{cb}|{\cal F}_{B\to D^*}(1)$ from
CLEO~\cite{CLEO}, the LEP experiments~\cite{LEP}, and
Belle~\cite{Belle}, our result implies
\begin{equation}
	10^3|V_{cb}| = \left\{
		\begin{array}{ll}
			45.9 \pm 2.4^{+1.8}_{-1.4} & \cite{CLEO} \\
			38.7 \pm 1.8^{+1.5}_{-1.2} & \cite{LEP} \\
			39.3 \pm 2.5^{+1.6}_{-1.2} & \cite{Belle} \\
		\end{array} \right. ,
\end{equation}
where the second, asymmetric error comes from adding all our
uncertainties in quadrature.
Here we have included the QED correction to ${\cal F}_{B\to D^*}(1)$
of~+0.007.

Since several groups have started partially unquenched lattice
calculations of spectrum and decay constants, we conclude with some
remarks on the prospects for~${\cal F}_{B\to D^*}(1)$.
In this context, ``partially quenched'' means that the valence and sea
quarks have different, and separately varied, masses.
The analysis presented here shows that the double ratios bring the
statistical precision under control, and that fitting the heavy-quark
mass dependence is straightforward.
Two of our larger systematic uncertainties will improve simply by
including dynamical quarks.
First, the self-consistent determination of the heavy-quark masses and
of $\alpha_s$ will improve.
At present, we believe the quenching bias in $\alpha_s$, which affects
the short-distance contribution, to be the largest source of uncertainty
from the quenched approximation.
Second, partially quenched numerical data are enough to extract the
physical result, because one can use the recently derived result in
partially quenched chiral perturbation theory~\cite{Savage:2001jw}.

The other two main sources of systematic uncertainty are the lattice
spacing dependence and the matching of lattice gauge theory to HQET
and~QCD.
The former is mostly a matter of computing.
Indeed, our present estimate may be conservative, as it is driven by
the coarsest lattice.
To decrease the matching uncertainties, one must calculate the
normalization factor to two loops and calculate the $1/m_Q^2$ corrections
to one loop.
The latter is not quite as hard as it might seem.
Heavy-quark symmetry protects the needed matrix elements, so one only
needs the one-loop calculation of the chromomagnetic term in the
effective Lagrangian (a $1/m_Q$ term) and the $1/m_Q$ and mixed $1/m_cm_b$
terms in the currents.
(An alternative to perturbation theory would be to develop a fully
non-perturbative matching scheme for heavy quarks, including the
$1/m_Q^n$ corrections.)

With the improvements from unquenched simulations,
a more detailed study of lattice spacing dependence,
and higher order matching calculations, it is conceivable that the error
on ${\cal F}_{B\to D^*}(1)$ could be brought to or below~1\%.
At this level, it would become crucial to compute, possibly by
similar methods, the slope and curvature of ${\cal F}_{B\to D^*}(w)$
near~$w=1$.
Then the determination of~$|V_{cb}|$ would not only become very precise,
but also truly model-independent.

\acknowledgments
We thank Aida El-Khadra for helpful discussions.
High-performance computing was carried out on ACPMAPS; we thank past
and present members of Fermilab's Computing Division for designing,
building, operating, and maintaining this supercomputer, thus making
this work possible.
Fermilab is operated by Universities Research Association Inc.,
under contract with the U.S.\ Department of Energy.
SH is supported in part by the Grants-in-Aid of the
Japanese Ministry of Education under contract No.~11740162.
ASK would like to thank the Aspen Center for Physics for hospitality
while writing part of this paper.

\end{document}